\documentclass[aps,prstper,twocolumn,showpacs,letterpaper,longbibliography,nofootinbib,floatfix]{revtex4-2}   

\usepackage{graphicx}
\usepackage[above,below]{placeins}	% allows use of \FloatBarrier command to force section barriers
\usepackage{times}
\usepackage{color}		% page numbers added later, when compiling the whole proceedings
\usepackage{enumitem}          % package for handling list formatting
\usepackage{float}
\usepackage{tabularx}
\newcolumntype{C}{>{\centering\arraybackslash}X}

\newenvironment{myquote}%
  {\list{}{\leftmargin=0.15in\rightmargin=0.15in}\item[]}%
  {\endlist}

\usepackage[dvipsnames]{xcolor}

\usepackage{appendix}

\begin{document}

\begin{titlepage}

  \title{Characterization of Upper-Level Undergraduate \\ Quantum Mechanics Courses in the U.S.}

  \author{Jesse Kruse}
  \affiliation{Department of Physics, University of Colorado, 390 UCB, Boulder, CO 80309} 
  \author{Molly Griston}
  \affiliation{Department of Physics, University of Colorado, 390 UCB, Boulder, CO 80309} 
  \author{Bethany R. Wilcox}
  \affiliation{Department of Physics, University of Colorado, 390 UCB, Boulder, CO 80309} 

   \keywords{quantum physics, quantum mechanics, upper-division, assessment, content survey, faculty perspectives}

  \begin{abstract}
    Upper-level, undergraduate quantum mechanics (QM) is widely considered a difficult subject with many varied approaches to teaching it and considerable variation in content coverage. For example, two common approaches to undergraduate QM instruction are spins-first, which focuses on the postulates of QM in spin systems before discussing wavefunctions, and wavefunctions-first, which focuses on the Schrödinger equation and its solutions for continuous functions in various potentials before discussing spin. These different approaches, along with the content variability in the textbooks used by instructors, may mean students learn different things in QM classes across the United States (U.S.). In this paper, we offer a characterization of QM courses based on survey responses from instructors at institutions across the U.S. With the responses of 76 instructors teaching QM courses (or sequences), we present results detailing their teaching methodologies, use of pedagogical resources, and coverage of QM topics. We find that the plurality of instructors in our sample are using traditional lecture, but many instructors are using interactive lecture or another non-traditional method. Additionally, instructors are using a wide variety of pedagogical tools (e.g., clicker questions). Many instructors in our sample reported teaching single-semester (or single-quarter) QM courses; these instructors report similar content coverage to instructors teaching first-semester (or quarter) QM courses, though their responses showed greater variability. We additionally report a comparison of content coverage between instructors using wavefunctions-first versus spins-first approaches, finding a large degree of overlap with differences in coverage for a few specific topics. These findings can help inform both future research and instructional efforts in QM education. 
  \end{abstract}

  \maketitle
\end{titlepage}

% INTRO
\section{\label{sec:intro}Introduction}

While Quantum Mechanics (QM) education has been a core part of the physics curriculum for many decades, it has recently been highlighted as an area of increased national attention due to the rise of Quantum Information Science (QIS). Governments around the world have funneled significant funding into developing QIS technologies such as novel quantum sensors with the potential to unlock new regimes of discovery, and quantum information processors that lay the foundation for realizing scalable quantum computation~\cite{dowling2003quantum, hughes2022assessing}. In the context of this increased attention, there is now a unique opportunity, and perhaps responsibility, for the physics education research (PER) community to reflect on the status of QM education at the undergraduate level to ensure we are preparing both a quantum-literate citizenry and a quantum-ready workforce to support the growing quantum industry \cite{fox2020preparing}. In this manuscript, we endeavor to characterize the current state of upper-level, undergraduate QM education in the United States (U.S.) as a first step toward enabling this reflection.

Previous studies have shown that QM education varies both in what is taught and the methods and perspectives used by instructors \cite{dubson2009faculty, siddiqui2017diverse}. One reflection of this variation is the variety of textbooks available for instruction \cite{griffiths2018introduction,mcintyre2022quantum,townsend2000modern,shankar2012principles,beck2012quantum}, though it has historically been unclear how many of these texts are widely used. Another reflection of this variation is the number of distinct QM research-based assessments (RBAs) that have been created to assess learning in these courses, each with different content and focuses \cite{mckagan2020physport}. There currently exist six upper-level QM RBAs \cite{cataloglu2002testing, sadaghiani2015quantum,marshman2019validation,singh2010surveying,goldhaber2009transforming, falk2004developing} covering topics that include wavefunctions, measurement, time dependence, probability, tunneling, and many other topics commonly seen in a first semester, upper-level QM course\footnote{As will be discussed in the Methods section, here we exclusively focus on upper-level quantum courses, which explicitly excludes lower and middle-division courses, such as Modern Physics, or dedicated Introductory QIS courses.} (QM1). However, there are many QM topics that are not covered by existing RBAs, including less ubiquitous QM1 topics (e.g., angular momentum, entanglement) and typical second-semester QM (QM2) topics. This poses an issue for physics education researchers whose goals are often to provide tools for instruction and assessment that will be applicable to as many courses as possible. 

Historically, both curriculum and assessment developers in PER have dealt with the issue of variable content coverage by focusing their efforts primarily on the core topics included in all or most courses in a particular core content area. However, a potentially more authentic approach could be to build curricular materials and assessments that accommodate the variety of topical areas covered in an instructor's course. The work reported here is part of a broader effort to fill this gap for quantum assessments, in particular. Specifically, we are engaged in creating a customizable QM assessment, called the Quantum Physics Assessment (QuPA), that will employ a flexible assessment design where instructors can choose which areas they want to assess and appropriate items will be provided algorithmically from a validated test bank. The goal of QuPA is to accommodate the natural variation in topical coverage in QM courses, including both QM1 and QM2 topics, and topics related to connected areas such as quantum computing.  
%This flexible assessment design and instructors' thoughts and opinions on it is the subject of previous work \cite{kruse_instructors_2024}. 
A crucial step in creating this new assessment is to understand which topics are actually covered in quantum courses and sequences to facilitate creation of an item bank with sufficient breadth. 

While the work reported here was motivated by, and embedded in, the development of QuPA, this paper focuses more broadly on characterizing the current state of upper-level, undergraduate QM courses in the U.S. This work will serve as a baseline for creating new assessment items that align with the priorities of instructors from a broad range of institutions. However, the potential impact of this work is not limited only to the development of RBAs. Indeed, we designed our methods and analysis to provide a comprehensive snapshot of QM education in the U.S. that will be useful both to inform instructors about common instructional practices (e.g., to promote appropriate expectations regarding the content knowledge of incoming graduate students) and development of curricular materials to target important topics without existing research-based curricula. Specifically, we utilized both interviews with faculty who have taught QM courses and faculty responses to a larger scale survey. In the interviews, faculty were asked to discuss their experiences teaching QM, including the topics covered, textbooks used, and pedagogical approaches. The results of the interviews informed the creation of the survey, which was distributed to physics departments across the U.S. The survey results provide detailed information about instructors' topical coverage, textbook usage, teaching methodologies, use of pedagogical resources, and more. 

The remainder of this paper is organized as follows. In Section~\ref{sec:lit}, we discuss the extensive literature around student learning in quantum mechanics and how it relates to and motivates this study. Following this, Section~\ref{sec:methods} outlines the methods used to design and analyze the interviews and survey, and outlines some of the limitations of this work. Section \ref{sec:results} lays out key findings and implications of this work. Finally, Section~\ref{sec:conclusion} discusses conclusions and future work.

% LIT REV /  BACKGROUND
\section{\label{sec:lit}Background}

There are two areas of existing research particularly relevant to the current work: studies investigating the scope of content coverage in quantum mechanics, and studies of student understanding or difficulties in quantum mechanics. 

\subsection{Scope of Quantum Mechanics Instruction}

Investigations of faculty consensus regarding content coverage in core physics courses has a long history in PER in the context of both curriculum \cite{chasteen2015educational} and assessment development \cite{rainey2020designing, meyer2024introductory}. These studies typically include a combination of interviews and survey responses from instructors teaching courses relevant to the content area in question. Indeed, there are several examples of these studies in the area of quantum mechanics that our work builds on. 

Perhaps the earliest reference to disagreement among faculty over the content of a QM1 course was provided by Dubson et al. \cite{dubson2009faculty} in 2009. In this work, the authors interviewed 27 faculty from 6 institutions regarding their teaching methodology and surveyed 20 QM textbooks. They found significant agreement on the list and order of topics but disagreement over the importance of certain key topical areas, such as the QM postulates, the treatment of measurement, and the physical interpretation of the wave function. While the results of this work suggest that historically there was significant consensus (at least as reflected by presentation in textbooks) in the topical coverage of QM courses, there are several reasons to suspect this finding may not hold through to the present day. Indeed, the landscape of traditional QM education has changed considerably in the past several decades. John Townsend, inspired by the work of Richard Feynman \cite{feynman1965lectures3} and J.J. Sakurai \cite{sakurai2020modern}, was the first to write an undergraduate quantum mechanics textbook that begins by introducing the quantum postulates using spin systems, rather than wave mechanics \cite{townsend2000modern}. In 2012, David McIntyre introduced another ``spins-first" quantum mechanics textbook targeting undergraduate students, which follows a similar progression of topics \cite{mcintyre2022quantum}. Certain institutions have reported adopting spins-first approaches; however, it is unclear how widespread this adoption is, as well as the prevalence with which each of the spins-first textbooks are used (see Sec.~\ref{sec:results}). This fundamental difference in how QM is taught (spins-first vs. wavefunctions-first) has sparked debate over which method is more effective at producing learning gains \cite{sadaghiani2016spin, riihiluoma2025sf}. Additionally, the explosion of QIS has introduced an entirely new area of focus to what might be taught or emphasized in QM courses \cite{meyer2024introductory}. 

Concerns about continued consensus on QM topics have been supported via more recent studies in this area. Siddiqui and Singh ~\cite{siddiqui2017diverse} made the claim in 2017 that ``there is no widespread agreement on the essential topics to teach in the college undergraduate QM course for physics majors, the order in which those topics should be taught, and the amount of time that should be spent on various topics." In this study, the authors surveyed twelve faculty members, six from the same institution and six from different institutions, about various aspects of teaching QM courses, including the goals of a QM course, the order of topics (e.g., wavefunctions vs spins first), and challenges faced when teaching QM. They found that QM instructors often had similar goals for an undergraduate QM course, but there were important differences of opinion on how to engage students, the order of topics, which topics they included in their course, and what level of emphasis should be placed on these topics. The results presented in this paper build on Siddiqui and Singh's ~\cite{siddiqui2017diverse} work by reporting results from a larger, more diverse respondent pool, as well as by asking additional questions regarding prerequisites, pedagogical resources used, and details of content coverage.

Another approach to analyzing the status of QM education was recently presented by Buzzell et al. \cite{buzzell2025quantum}. In this work, they analyzed the course catalogs of 188 U.S. research intensive institutions of higher education to assess the role of QM courses in the undergraduate curriculum. The institutions were selected from the US News Rankings of ``The Best Physics Programs" \cite{usnews_physics_rankings_2023}. They found every institution surveyed required at least one course on quantum concepts, 92\% required two courses, and 50\% required three courses. They also analyzed the topics taught in the quantum curriculum and found that the Schrödinger equation and QM in 3D were the two most common themes (overarching topical categories in their coding scheme) with a distribution of coverage for other themes. In addition, the authors of this study found that around 74\% of instructors used a wavefunctions-first approach rather than a spins-first approach. The prevalence of the wavefunctions-first approach is consistent with the findings of this paper (See Sec.~\ref{sec:results}). 

Our work complements the work of Buzzell et al. by addressing the question of content coverage in QM courses with a different methodological approach and a finer grain size in terms of topical coverage. For a subset of the courses in their data set, Buzzell et al. utilized analysis of course catalogs and syllabi to determine which topics were covered in those courses, whereas our approach directly compiles instructors' perspectives on their teaching of upper-level QM courses via interviews and surveys. Both of these approaches have methodological limitations such as selection effects (e.g., instructors select whether to respond to a survey or interview request) and data fidelity (e.g., not all instructors include enough information in a syllabus to allow for fine grained information into their content coverage); however, we believe that together, these two studies have the potential to provide a more comprehensive view on QM education.\footnote{Note that Buzzell et al. included Modern Physics courses amongst their data set whereas we focus only on upper-division QM courses beyond the level of modern physics.} Throughout our analysis and results, we will provide comparisons between our findings and those of Buzzell et al.

\subsection{Student Learning of Quantum Mechanics}

Significant research has been done to understand student reasoning and difficulties in learning QM \cite{singh2015review}. Emphasis on this subject is, in part, because QM requires students to undergo a ``paradigm shift" away from the classical phenomena that students see up until this course. Working with probabilistic observables that sometimes cannot be simultaneously observed to arbitrary precision can be difficult for students to understand when coming from a deterministic viewpoint \cite{singh2015review}. In addition, there are many non-intuitive quantum topics and phenomena like tunneling \cite{wittmann2005addressing}, wave-particle duality \cite{ambrose1999investigation}, superposition of states \cite{singh2008student}, and collapse of the wavefunction \cite{singh2001student} that have historically provided difficulties for students. 

In light of these student difficulties and barriers to learning, physics education researchers have created many research-based curricular materials for improving QM education such as online and in-person tutorials \cite{singh2008interactive, emigh2020based, brown2015developing, corsiglia_effectiveness_2022} and dynamic simulations \cite{mckagan2008developing, kohnle2017sim}. These materials have been shown to improve students' understanding of quantum concepts despite the difficulties associated with a novel paradigm of thinking \cite{brown2015developing, keebaugh2018developing}. 

Although the content of this paper does not directly contribute to the body of work studying student difficulties, it does provide more detailed information about how QM is being taught, including the degree to which various topics are covered in a broad sampling of courses. Existing studies on student difficulties in QM have been motivated by the previous work identifying topical coverage (e.g., \cite{riihiluoma2025sf}). Given the aforementioned changes in QM instruction, the updated and expanded information provided in this paper can help guide physics education researchers and curriculum developers in prioritizing the development of necessary materials and identifying areas where more research on student difficulties may be warranted.

\section{\label{sec:methods}Methods}

In this study, we employed an iterative, mixed methods research design that consisted of interviews followed by a broadly distributed survey. In this section, we will articulate the relevant portions of the interview and survey design, respondent demographics, and limitations of these methodologies. 

\subsection{Interview methodology}

To solicit participants, we compiled contact information for physics departments in the U.S. that offered a QM course (N = 470). No institutions were excluded unless we could not find contact information available online or they explicitly did not require any QM courses in their curriculum. In our email solicitation, we asked for faculty who had taught: modern physics; graduate-level QM; graduate or undergraduate quantum information or quantum computation; and/or upper-level undergraduate QM to participate.\footnote{While this analysis is focused on undergraduate QM, we initially solicited participation from faculty members who had taught courses that may have some overlap with undergraduate QM. Including such faculty in our interviews ensured we were not overly restrictive in our scope. The survey results, however, are focused specifically on undergraduate QM.} Overall, 35 faculty volunteered to participate in interviews; however, one interviewee was excluded from the analysis as they did not answer a sufficient number of questions. Of our participants, 28 reported having taught upper-level undergraduate QM, 4 had taught modern physics, 2 had taught graduate QM, and 2 had taught a quantum information course (2 instructors had taught two different courses). The 34 interview participants represented faculty from 33 unique institutions. Information about the institutions represented can be found in Table \ref{tab:interview_institutions}. 

\begin{table}[t!]
    \centering
    \caption{Carnegie classification \cite{Carnegie:2021}, control (public/private), and minority-serving institution (MSI) status for each of the 33 institutions represented by the 34 interview participants. MSI status was determined using the NASA 2024-2025 List of Minority Serving Institutions (HSI: Hispanic-Serving Institution, AANAPISI: Asian American and Native American Pacific Islander-Serving Institution, HBCU: Historically Black College or University) \cite{NASA:2024}. *These categories are not mutually exclusive.}
    \begin{tabular}{l c @{\hspace{15pt}} l c @{\hspace{15pt}} l c}
         \hline \hline Carnegie classification &  & Control &  & MSI Status &  \\
         \hline Doctoral (R1) & 6 & Public & 21 & HSI & 5 \\
         Doctoral (R2) & 6 & Private & 12 & AANAPISI & 2 \\
         Doctoral/Professional & 2 &  &  & HBCU & 1 \\
         Masters (M1) & 10 &  &  &  &  \\
         Masters (M2) & 1 &  &  &  & \\
         Baccalaureate & 8 &  &  &  & \\[6pt]
         Total & 33 & Total & 33 & Total & 7*\\
         \hline \hline
    \end{tabular}
    \label{tab:interview_institutions}
\end{table}

The interviews were semi-structured and each lasted approximately an hour. They consisted of two primary components. The first component, focusing on the instructor's interaction with an assessment creation interface, was the subject of a previous publication \cite{kruse2024instructors}. The second component targeted the instructor's experience teaching QM and asked them to identify what from a list of QM topics they would teach and assess in their class. This list of topics was initially generated by the authors but was updated and added to by instructors in each interview. The interview data were then used to inform the design of a corresponding survey. 

\begin{table*}[t]
    \centering
    \caption{Demographics of instructor survey respondents. *Respondents could select multiple options.}
    \begin{tabular}{l c @{\hspace{15pt}} l c @{\hspace{15pt}} l c @{\hspace{15pt}} l c}
         \hline \hline Race &  & Gender & & Position & & Time Teaching \\
         \hline White & 60 & Male & 58  & Professor & 44 & 21+ years & 40\\
         Asian & 10 & Female & 11 & Associate Professor & 20 & 16-20 years & 9\\
         Prefer not to say & 4 & Prefer not to say & 4 & Assistant Professor & 8 & 11-15 years & 13\\
         Hispanic or Latinx & 2 & Non-binary/third-gender & 2 & Other & 3 & 6-10 years &  6\\
         Did not respond & 1 & Did not respond & 1 & Lecturer & 1 & 0-5 years & 8\\[6pt]
         Total & 76* & Total & 76* & Total & 76 & Total & 76\\
         \hline \hline
    \end{tabular}
    \label{tab:survey_dem}
\end{table*}

\subsection{Survey Design}\label{sec:surveymethods}

The purpose of the survey was to solicit information from a broader array of faculty at U.S. institutions offering at least one upper-level, undergraduate QM course. Specifically, we aimed to identify instructors' teaching methodology and the topical areas they consider important, peripheral, or extraneous. The complete survey, with all available responses, can be found in Appendix \ref{app:survey}. Here, we will cover the aspects of the survey structure necessary to understand and interpret the results. The survey consists of four primary categories: institutional information, course-level data, topical coverage, and optional demographics.   

\emph{Institutional information}: Instructors were asked if their institution follows a semester or quarter system; this question was necessary to frame later questions. They were also asked which QM courses their institution offers. Of the 76 instructors who responded to our survey, 68 taught at semester-based institutions and 8 taught at quarter-based institutions, suggesting that our results are primarily reflective of semester-based institutions. 

\emph{Course-level data}: Instructors were first asked which upper-level undergraduate QM courses they had taught in the previous 5 years, with slightly different options available for instructors at semester and quarter-based institutions. Instructors had the option to select multiple courses; however, they only filled out the rest of the survey survey once (i.e., for instructors who had taught multiple courses, their responses are not differentiated by course). While this presents certain limitations, which will be discussed below, our priority was to limit survey fatigue and maximize the number of complete responses. Instructors were then asked a series of questions about the course(s) they had selected, regarding formal prerequisites, textbook usage/order of content coverage, teaching style, class size, and changes in teaching due to the Covid-19 pandemic. 

\emph{Topical coverage}: The instructors were then asked about the topical coverage in the QM courses they had taught. They selected topics from a list of 84 topics that were generated via textbook review and revised by instructors in the interviews. The 84 topics were split into five pages to reduce cognitive load. For each of these topics, we presented them with four response options for their level of coverage: \textit{important}, \textit{peripheral}, \textit{assumed prior knowledge}, and \textit{do not cover/use}.\footnote{The use of four levels of coverage was inspired by Meyer et al.~\cite{meyer2024introductory}.} At the beginning of each page, we provided them with the following operational definition of these categories:

\begin{itemize}
    \item \textit{Important}: You teach this topic in depth and assess students' comprehension and retention of it.
    \item \textit{Peripheral}: You teach this topic but not in depth. That could mean mentioning it in passing or covering it for part of a lecture. It is also something not directly assessed.
    \item \textit{Assumed Prior Knowledge}: You do not teach this topic, but you use it, assuming prior knowledge or exposure from your students.
    \item \textit{Do Not Cover/Use}: You do not teach or use this topic in your class.
\end{itemize}

\noindent Respondents were also provided an opportunity to report any topics missing from our list. 

It is worth reflecting on the overlap between the topics generated by Buzzell et al. \cite{buzzell2025quantum} and those included in our list. As our list was designed to be as comprehensive as possible while theirs was generated from thematic analysis of syllabi, we do not expect complete agreement between the lists. Indeed, we expect that our list will include a number of less common topics that likely would not appear in analysis of syllabi; however, it is valuable to note the degree of overlap, particularly with respect to any topics that might be missing from our list. Buzzell et al. \cite{buzzell2025quantum} generated a list of 59 topics spanning both modern physics and quantum mechanics. In comparison with our list of 84 topics (see Sec.~\ref{sec:results}) generated through textbook analysis and faculty interviews, there was a significant degree of overlap. Aside from specifically modern physics topics (e.g., the photoelectric effect), almost all topics included on the Buzzell et al. list appeared either verbatim on our list (e.g., harmonic oscillator) or were implicitly included in our list of topics (e.g., our list did not include ``spontaneous emission" but did include ``Einstein coefficients" and "selection rules"). We identify three topics on the Buzzell et al. list that do not appear on our list in any form: neutrino oscillations, the adiabatic theorem, and Berry's phase. However, neutrino oscillations and Berry's phase were both supplied by survey respondents as additional topics in our optional 'Other' box.\footnote{We note that Berry's phase is typically introduced with the adiabatic theorem \cite{griffiths2018introduction}.}

\emph{Optional demographics}: The final section of the survey asked optional demographic questions about the instructor, as well as information about their institution. 

\subsection{Participant Demographics}

We collected a total of 76 complete survey submissions. The demographics of the respondents can be found in Table \ref{tab:survey_dem}. Respondents were given an opportunity to provide the name of their institution, and 75 of the respondents provided one, while one respondent instead provided only institutional information. The 76 instructors represented 74 unique institutions, where we have considered the unspecified institution to be a unique institution for reporting purposes. Information about the represented institutions can be found in Table \ref{tab:survey_institutions}. Note that relative to Buzzell et al. \cite{buzzell2025quantum}, our data set has a smaller fraction of institutions with very high research activity, a smaller fraction of public institutions, and a slightly higher fraction of minority serving institutions (MSIs). This suggests that the differing methodologies used in these two studies has indeed resulted in different selection effects. 

In Table \ref{tab:semester_instructors}, we provide a breakdown of the courses respondents in our sample had taught within the last 5 years. For our analysis, it is important to distinguish between instructors who taught the first course of a multi-course sequence and those who taught a stand-alone course. To address the concern that some instructors might misinterpret the intended distinction between QM1 and QM (single-semester or single-quarter), we cross-checked their responses to the questions asking which QM courses they had taught and which QM courses their institution offers. For any potential inconsistencies (e.g., an instructor stating they had taught QM1 but that their institution only offers a single-semester QM course), we verified the provided information with the institution's course catalog. Based on this verification process, we re-classified three instructors. Table \ref{tab:semester_instructors} shows the final, corrected classifications. We note that the majority of survey respondents had recently taught a single-semester QM course, suggesting that a large fraction of the represented physics programs (on the order of half) offer only one semester (or 1-2 quarters) of QM instruction. The next most represented course taught was QM1 of a two semester sequence and then both QM1 and QM2. 

\begin{table}[t]
    \centering
    \caption{Carnegie classification \cite{Carnegie:2021}, control (public/private), and MSI status for each of the 74 institutions represented by the 76 survey respondents. MSI status was determined using the NASA 2024-2025 List of Minority Serving Institutions (HSI: Hispanic-Serving Institution, AANAPISI: Asian American and Native American Pacific Islander-Serving Institution, PBI: Primarily Black Institution, ANNH: Alaska Native- and Native Hawaiian-Serving Institution) \cite{NASA:2024}. One instructor did not provide the name of their institution but self-reported institutional information. For all other institutions, the information was determined by the authors using the institution name. *These categories are not mutually exclusive.}
    \begin{tabular}{l c @{\hspace{15pt}} l c @{\hspace{15pt}} l c}
         \hline \hline Carnegie classification &  & Control &  & MSI Status &  \\
         \hline Doctoral (R1) & 17 & Public & 42 & HSI & 11 \\
         Doctoral (R2) & 11 & Private & 32 & AANAPISI & 10 \\
         Doctoral/Professional & 4 &  &  & PBI & 1 \\
         Masters (M1) & 17 &  &  & ANNH & 1 \\
         Masters (M2) & 2 &  &  &  & \\
         Masters (M3) & 6 &  &  &  & \\
         Baccalaureate & 17 &  &  &  & \\[6pt]
         Total & 74 & Total & 74 & Total & 16*\\
         \hline \hline
    \end{tabular}
    \label{tab:survey_institutions}
\end{table}

\begin{table}[b]
    \centering
    \caption{Number of instructors at semester-based and quarter-based institutions that taught each course in the last five years.}
    \label{tab:semester_instructors}
    \begin{tabular}{l|c}
         \hline \hline Semester Course(s) Taught & Number of Instructors \\
         \hline QM (Single Semester) & 39 \\
         QM 1 & 14 \\
         QM 1 and QM 2 & 14 \\
         QM 2 & 1 \\
         \hline Quarter Course(s) Taught & \\
         \hline 
         QM (Single Quarter) & 1 \\
         QM 1 and QM 2 & 4 \\
         QM 2 & 1 \\
         QM 3 & 2 \\
         \hline
    \end{tabular}

\end{table}

\subsection{Analysis}

In our data collection, we made the distinction between quarter-based and semester-based institutions. For our topical analysis, however, it is useful to combine the two groups. Where relevant, we have included QM1 from quarter-based institutions with QM1 from semester-based institutions. Similarly, we have included QM2 and QM3 from quarter-based institutions with QM2 from semester-based institutions. This approach has limitations, notably that QM2 from quarter-based institutions potentially shares content with QM1 from semester-based institutions. Part of the rationale for our groupings is that at least one instructor at a quarter-based institution noted that their institution requires a first quarter QM course but does not require a second quarter QM course. While not perfect, this grouping allows for the inclusion of quarter-based institutions in our analysis. Details regarding methodological choices for specific elements of analysis appear in the relevant section.

Throughout our results, we present raw counts or percentages from survey responses. As our goal is focused on understanding broader trends rather than making specific quantitative claims, we do not perform specific statistical tests to investigate statistical significance of, for example, differences between the frequencies of different topical areas. In some cases, we discuss comparisons of these general trends between different types of courses (notably spins-first vs. wave-functions first); however, these should be interpreted as qualitative rather than quantitative comparisons.  

\subsection{Limitations}

There are several important limitations to consider in interpreting the results of this analysis. First, this study relies on both interview and survey volunteers, which may over-represent faculty from institutions where the research or teaching loads are more likely to provide space in faculty schedules to respond to such requests. Similarly, interview volunteers may also self select for instructors who are more likely to know about and utilize active learning techniques. Additionally, our data focus exclusively on the faculty perspective (rather than the student perspective) regarding the content of undergraduate QM course. This was an intentional choice because undergraduates likely do not have sufficient experience with the topic of QM. However, we note that the student perspective is an important element of any educational reform, and encourage any researcher or instructor utilizing our results to consider the student experience and perspective. 

Additionally, we designed our survey to accommodate instructors who have taught any combination of a single semester QM course, a sequence of QM courses, or only one course within a sequence of QM courses in the last 5 years. However, instructors who had taught multiple courses were not asked to submit separate responses for each course in order to reduce survey fatigue. This introduced several constraints, including that we cannot distinguish where topics were taught for instructors who had taught multiple courses (i.e., we have their full topical focus but not which course introduced it). We have discussed our approach for dealing with this constraint in our discussion of topical coverage in Sec. \ref{sec:topic_coverage}.

% RESULTS
\section{\label{sec:results}Results}

In this section, we first present and discuss results regarding reported prerequisites, textbooks, teaching methodology, and pedagogical resources used by instructors. Then, we discuss the coverage of various QM topics. 

\subsection{Course-level characteristics}

\begin{figure}
    \centering
    \includegraphics[width=\linewidth]{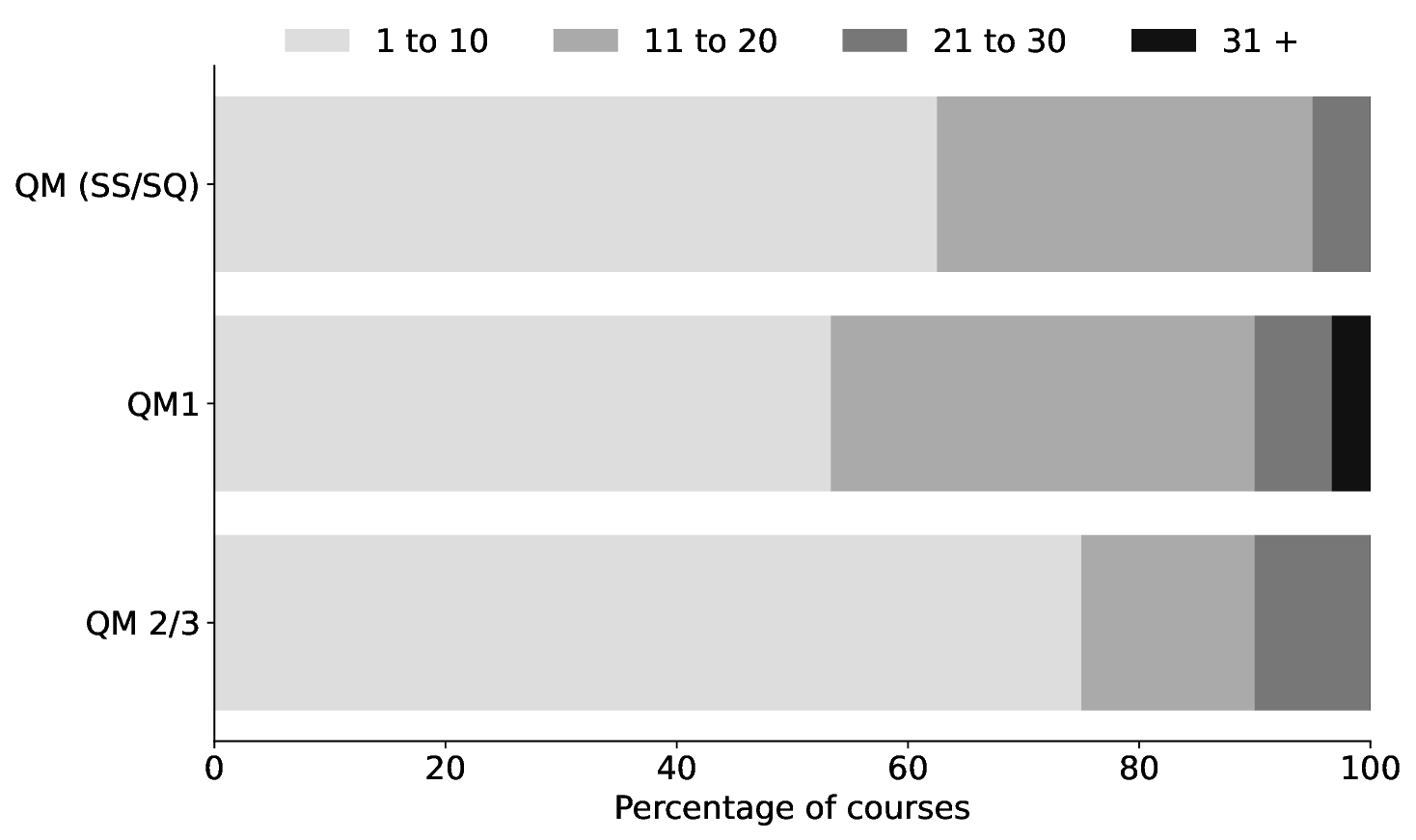}
    \caption{Distribution of class sizes for courses taught by survey respondents. Quarter-based and semester-based institutions are grouped together. Two instructors did not provide class size information, so their courses are not included in this plot.}
    \label{fig:size}
\end{figure}

Here, we report the distribution of class size, as well as the required prerequisites. The question asking instructors for their average class size was open-ended, so we have grouped responses into bin sizes of 10. If an instructor reported a range, the average of the range was used for classification. Instructors who reported having taught multiple courses were requested to specify the course size for each class. Figure \ref{fig:size} shows the distribution of class size, split by the type of course. We see that most courses represented in our sample are small, with 10 or fewer students. 

Information on prerequisites for the courses included in our data set were collected in two ways. The survey explicitly asked instructors to report on the prerequisites for their courses; however, after initial analysis, it became clear that the question of prerequisites was interpreted in different ways by different respondents. Some respondents appeared to list only the immediate prerequisites for their course, while other listed  all courses that are implicitly prerequisites. For example, some institutions require Classical Mechanics 1 or 2 as a prerequisite for QM and list ordinary differential equations as a pre- or co-requisite for Classical Mechanics. Thus implicitly, ordinary differential equations becomes a prerequisite for QM but one that is not always explicitly listed. Some respondents seemed to take into account these implicit prerequisites while others did not. 

\begin{figure}
    \centering
    \includegraphics[width=\linewidth]{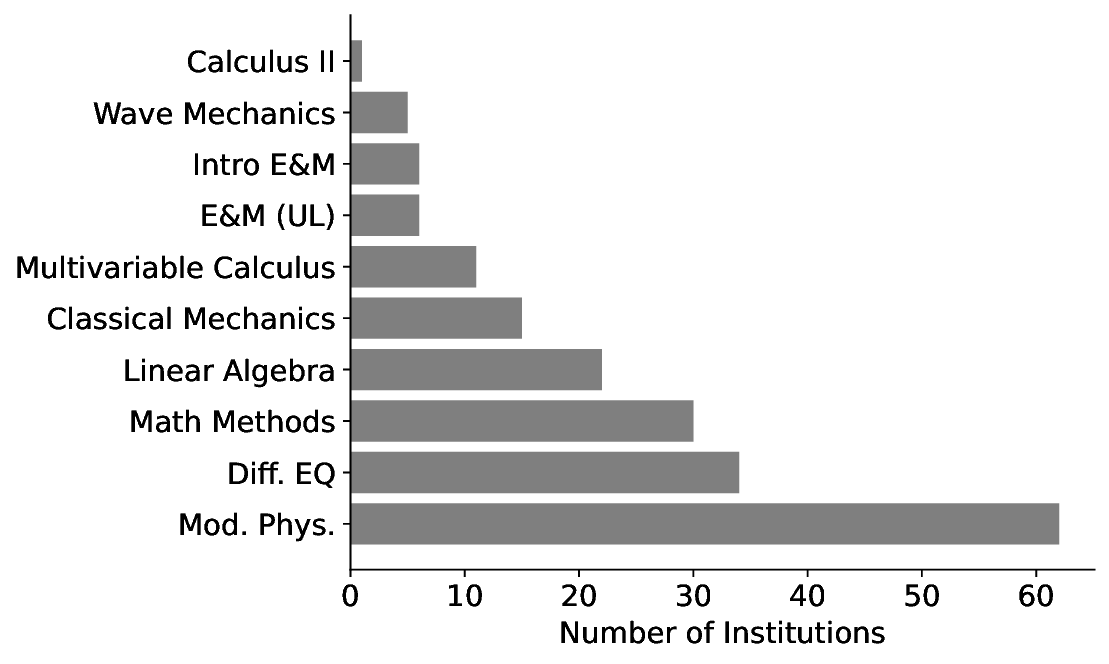}
    \caption{Distribution of explicitly listed prerequisites for QM1 and single-semester/quarter QM courses across all institutions in the sample (N=74). Counts were derived by analysis of course catalogs for each institution in the study. (Mod. Phys. = Modern physics/introduction to QM, UL = Upper level, E\&M = Electricity and Magnetism, Diff. EQ = Differential equations)}
    \label{fig:qm_prereqs}
\end{figure}

To address this limitation and ensure uniformity, we utilized institutional information collected from the survey to pull online course catalogs for each institution to identify the explicitly listed prerequisites for their QM course(s). When we look at the listed prerequisites on course catalogs for QM1 and single-semester/quarter QM courses, we find the distribution shown in Figure \ref{fig:qm_prereqs}. We see that modern physics is the most common prerequisite followed by ordinary differential equations, mathematical methods for physicists, and linear algebra.\footnote{The ordering of the prevalence of prerequisites is almost identical to what was reported by respondents, except that according to the course catalogs, math methods is more common than linear algebra as a prerequisite for QM.}  It is interesting to note that not every institution required a named ``modern physics" course to take upper-level QM. In these cases, some institutions instead required a wave mechanics course or intermediate classical mechanics course. For the 30 institutions that offered a QM2 or QM3 course, the only prerequisite listed for each was the prior QM course. While Buzzell et al. \cite{buzzell2025quantum} noted a tendency in their sample of courses not to require linear algebra prior to a Modern Physics course, for the upper-division courses that are the focus of our study, it is apparent that differential equations, math methods, and linear algebra are the most common math requirements for these courses. 

\subsection{QM Textbooks}

\begin{figure}[b]
    \centering    \includegraphics[width=\linewidth]{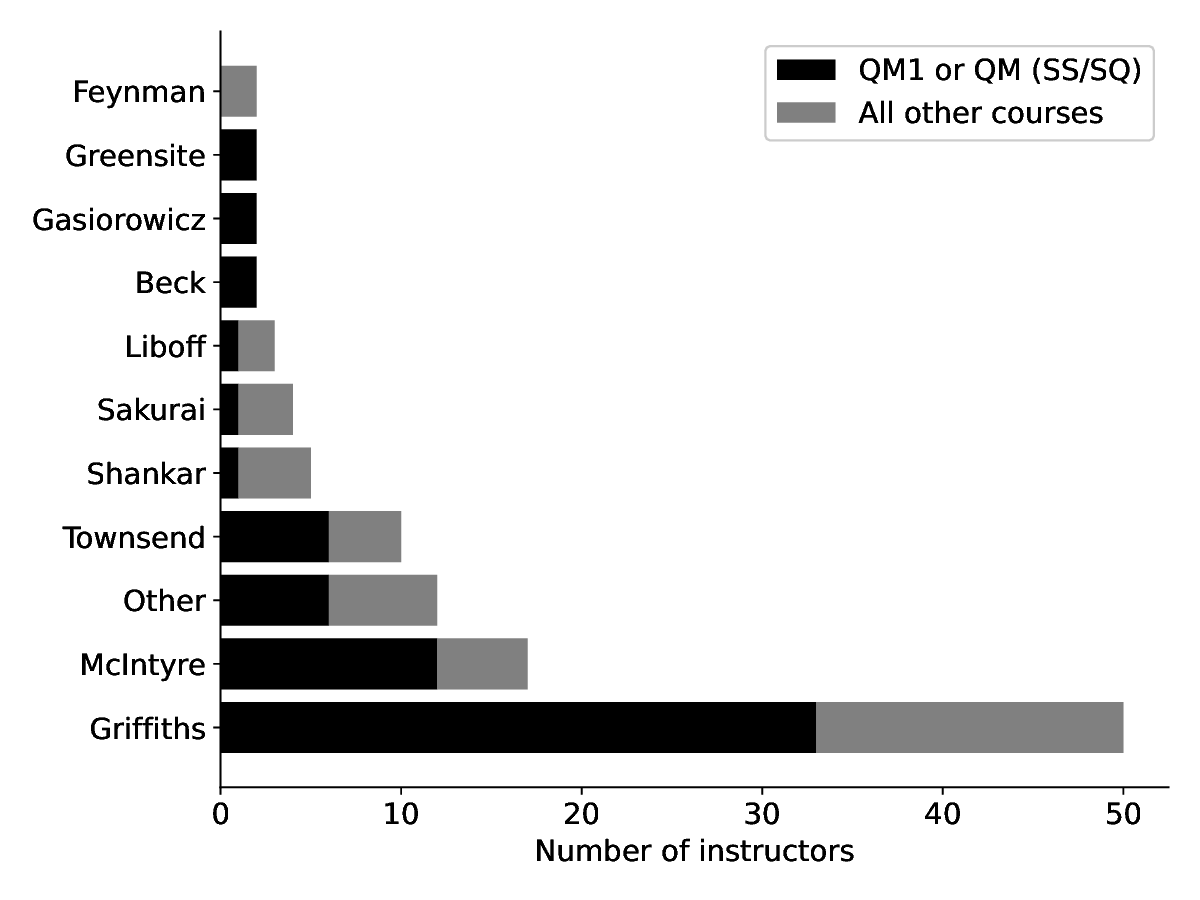}
    \caption{Distribution of textbooks used by surveyed instructors. Textbooks are included in the ``Other" category if they were only used by one instructors (i.e., this category includes 12 different textbooks). The textbooks are labeled by the first author and different editions of textbooks have been collapsed. The entire bar shows the total count from all instructors, and the black portion of the bar shows the contribution from instructors who had only taught a QM1 or single semester/quarter QM course.}
    \label{fig:textbooks_all}
\end{figure}

The majority (71 of 76) of instructors stated they follow a textbook, while 4 stated they used only lecture notes, and 1 said they do not use either a textbook or lecture notes. As shown in Figure \ref{fig:textbooks_all}, Griffiths QM \cite{griffiths2018introduction} is the most used textbook, followed by McIntyre \cite{mcintyre2022quantum} and then Townsend \cite{townsend2000modern}. However, there are a number of other texts used by multiple instructors (see Fig.~\ref{fig:textbooks_all}) along with 11 other texts reported by only one respondent each (see Appendix \ref{app:details}). This variation in textbook use shows up even when restricting our view to just single-semester or QM1 courses (see Fig.~\ref{fig:textbooks_all}), suggesting that the variation is not a result of a preference for different texts when including more advanced QM content. 

The prevalence of Griffiths suggests that a wavefunctions-first approach to QM instruction is still the most common approach (roughly two-thirds of respondents). This is consistent with the findings of Buzzell et al. via their analysis of syllabi \cite{buzzell2025quantum}. That said, counting both McIntyre and Townsend as spins-first approaches, we also find a significant fraction of courses (roughly a third) utilizing some version of this approach. There is also a long tail of other texts used either in conjunction with a more prevalent text (i.e., one of the texts in Figure~\ref{fig:textbooks_all}) or as the main text for the course.

Since content focus, notational conventions, and consistency of notations such as Dirac notation can vary significantly depending on the text uses, this finding has important implications for both assessment and curricular development. Variations in content coverage and focus emphasize the potential value of flexible assessments that can be customized to ensure alignment with the instructor's priorities. It also suggests that modular curricular materials that can be slotted into courses on an as-needed basis would represent a significant resource to the instructional community. 

\subsection{Teaching Methodology}

\begin{figure}[b]
    \centering
    \includegraphics[width=\linewidth]{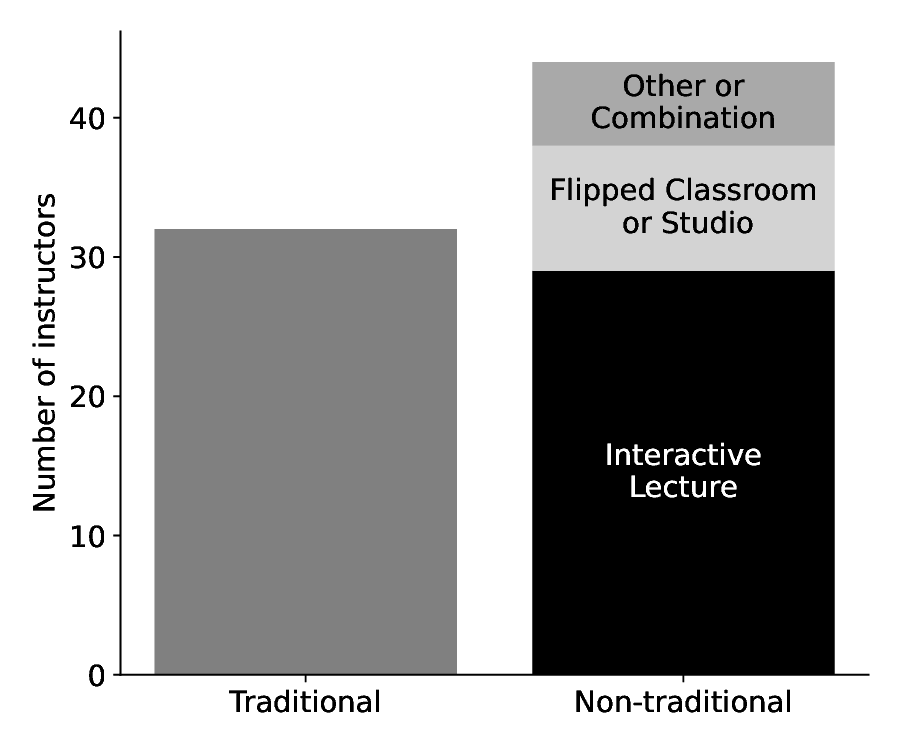}
    \caption{Number of instructors using each teaching style (single response only). Traditional lecture is defined as lecturing most of the class with notes on a powerpoint or written on a board and students ask questions of their own volition. Interactive lecture is defined as lecturing part of the class period, using concept questions or group discussion or tutorials and student activities. Flipped classroom/studio is defined as minimal lecturing with class period driven by students' questions or work on homework or projects.}
    \label{fig:teaching_styles}
\end{figure}

Our survey also asked participants about their teaching methodology in their classroom. The options they could pick from included: traditional lecture, interactive lecture, flipped classroom/studio, or an `other' option where instructors could provide their own description.\footnote{This question included operational definitions of each of these approaches to ensure respondents had a consistent interpretation. } As shown in Fig.~\ref{fig:teaching_styles}, the two leading methods were traditional lecturing and interactive lecture, with flipped classroom coming in last. Respondents who selected the `other' response described a mixture between these teaching styles.

We followed this question with an optional free-response text box for instructors to explain their motivations for using this teaching style. These responses were analyzed to identify common themes among the stated reasons for their chosen teaching style. Here, we will present example quotations corresponding to each identified theme. For traditional lecture (which represented ~40\% of respondents), some instructors expressed the idea that traditional lecture was simply the most effective. For example,

\begin{myquote}
    \textbf{P6:} \textit{It seems to work the best.} \\
    \textbf{P10:} \textit{I feel that lectures combined with problem sets are effective at teaching this material.} 
\end{myquote}

\noindent The majority of other justifications for traditional lecture were logistical in nature. Some instructors noted the difficulty inherent in other approaches:

\begin{myquote}
    \textbf{P2:} \textit{In general, I think that the flipped scheme has more pedagogical justification and better results. It's just more difficult to do, particularly starting out in a new prep.} \\
    \textbf{P25:} \textit{This is my first time teaching this class, and it's easier to prepare standard lecture.}
\end{myquote}

Other instructors expressed concerns about the time required for more active methods and the corresponding reduction in content coverage:

\begin{myquote}
    \textbf{P7:} \textit{It is the most efficient way of delivering the knowledge.} \\
    \textbf{P17:} \textit{There is too much material to cover to use the other methods.} 
\end{myquote}

\noindent Finally, several instructors noted a lack of institutional support to run their course in any other format than traditional lecture:

\begin{myquote}
    \textbf{P46:} \textit{No support to run the course in any other fashion.} \\
    \textbf{P49:} \textit{There are very little resources to apply to courses at this level and our teaching loads have become too high to spend too much time designing custom resources for interactive lectures...}
\end{myquote}

These justifications for a traditional teaching style over other research-based methods are consistent with prior work on instructional change  \cite{henderson2007barriers,henderson2011facilitating}. Specifically these findings are consistent with Henderson and Dancy's \cite{henderson2007barriers} finding that instructors often point to situational factors such as lack of time or institutional support to justify using traditional lecture.

Alternatively, some faculty argued for the importance of interactive engagement in these course. For example, some stated that they perceived better student engagement with the material:

\begin{myquote} 
    \textbf{P5:} \textit{It seems to strike the right balance for many of my students between being actively engaged and getting what they are looking for from me.} \\
    \textbf{P61:} \textit{Better learning, more engagement, much more fun.} 
\end{myquote}

\noindent Additionally some instructors stated that interactive lecture was more effective at revealing difficulties with concepts and procedures:

\begin{myquote}
    \textbf{P37:} \textit{I think that it leads to better learning and I can definitely see, in real time, where students are having difficulties.} \\
    \textbf{P42:} \textit{The students learn better and develop a sense of community when doing group work.} \\
    \textbf{P59:} \textit{I find the students understand a lot of the material once they get a chance to discuss the topic.} 
\end{myquote}

\noindent Difficulties with other methods of instruction were also cited as a reason for using interactive lecture:

\begin{myquote}
    \textbf{P9:} \textit{More interesting for students and instructors than traditional lecture.} \\
    \textbf{P23:} \textit{I have done the flipped classroom before, but students have increasingly struggled with that method and so I moved more toward a mix of lecturing and active discussion.} 
\end{myquote}

Lastly, some instructors cited physics education research as a primary reason for using interactive lecture in their classroom:

\begin{myquote}
    \textbf{P43:} \textit{Physics education research has clearly shown that students learn more from interactive engagement in the classroom than from traditional lecture, so I use an interactive approach in all of my courses.} \\
    \textbf{P58:} \textit{Results of PER groups that indicate that this style of instruction leads to better conceptual understanding...} 
\end{myquote}

For the flipped classroom style of teaching, instructors cited effectiveness as a primary reason for use:

\begin{myquote}
    \textbf{P18:} \textit{Since there are other good avenues for content delivery ... our time together in the same space is best spent through interactive ... intentional practice --- primarily working through examples, problems, and discussion questions.} \\
    \textbf{P74:} \textit{My motivation is that students need to be actually working with the concepts to learn them.}
\end{myquote}

\noindent Some of the instructors who used the flipped classroom methodology stated that they experienced better student engagement in this format:

\begin{myquote}
    \textbf{P16:} \textit{Maximizing time to address student questions and prevent students from spending a lot of time out of class being stuck on either a math ``trick" or a conceptual misunderstanding.} \\
    \textbf{P75:} \textit{I've found that this is better than the traditional method in at least one way: the students end up doing much more computation and practice than they otherwise would just reading the book outside of class and ``passively" listening to lecture during class time.} 
\end{myquote}

Finally, some of the instructors stated that they used a flipped classroom approach because a small class size allowed for it:

\begin{myquote}
    \textbf{P66:} \textit{This is my preferred style. Class is small, too (N=4).} \\
    \textbf{P73:} \textit{Small classes allow it. The availability of online solutions to homework problems has driven me to having students work out problems in class.}
\end{myquote}

\subsection{Pedagogical Resource Use}

After asking the instructors which teaching style they used and their motivations, instructors were asked to select the pedagogical resources/approaches they used in their teaching. Table \ref{tab:tools} shows the distribution of responses. We see that the top three most used were in-class group work, pre-class readings, and simulations. Responses in the Other category were primarily clarifications of approaches that combined different elements of multiple options. The only novel element noted in the Other responses was the inclusion of computational projects or assignments, which was given by 4 respondents.

\begin{table}[h]
\caption{Distribution of instructors who used each pedagogical research (N=76). This was presented as a list from which respondents could select multiple items.}
\label{tab:tools}
\begin{tabular}
{l @{\hspace{15pt}} c}
\hline \hline
Tool &  Count (out of 76)\\
\hline
In-Class Group Work & 43 \\
Pre-Class Readings & 39 \\
Simulations & 36 \\
HW Corrections & 27 \\
Exam Corrections & 25 \\
Concept/Clicker Questions & 24 \\
Tutorials & 23 \\
Final Papers or Projects & 18 \\
Outside Group Work & 16 \\
Other & 13 \\
Research-based Assessments & 12 \\
Demos & 9 \\
Pre-Class Quizzes & 7 \\
Group Exams & 7 \\
Online Tutorials & 5 \\
Oral Exams & 4 \\
\hline
\end{tabular}
\end{table}

It is notable that research-based assessments (RBAs) were used by only 12 of the 76 instructors who participated in this survey. There could be many reasons for this relatively low uptake of RBAs. For example, this result could suggest that many QM instructors are unaware of RBAs as a potential tool. Alternatively, it could be that many QM instructors do not find the existing RBAs for QM valuable. Either way, this suggests that assessment developers likely need to consider how to both increase awareness of these tools and ensure that they are meeting the needs of QM instructors broadly. Indeed, as discussed in Sec.~\ref{sec:intro}, this work was motivated, in part, to support efforts to develop a new QM assessment featuring flexible content coverage with the goal of ensuring the new instrument can accommodate a wide range of courses, instructional styles, and content priorities.

\subsection{QM Topic Coverage}\label{sec:topic_coverage}

The next section of the survey asked each instructor to rate their level of coverage for a list of 84 topics that could be taught and assessed in an upper-level, undergraduate QM course. The full data set is presented in Appendix \ref{app:topic}. Here, we report the responses by class type to clarify content coverage for each type of course. We first consider QM1 and QM2 courses and then consider single-semester or single-quarter QM courses. While one could make the case for including single-semester QM courses with QM1 courses, we consider it important to first analyze single-semester courses separately. This way, we can assess to what extent single-semester courses typically cover only QM1 content or if they cover a combination of QM1 and QM2 content. 

Throughout this section, we report specific percentages based on instructor responses to our survey. However, we advise the reader to keep in mind the sample sizes (provided for each Table) as they interpret the percentages. For example, topics covered by 100\% of QM1 instructors in our sample are certainly not all covered by 100\% of QM1 courses in the country. The topic lists we report are intended to serve as a guide when considering topical coverage in QM courses, but they should be interpreted with the appropriate context. 

\begin{figure}[b]
    \centering
    \includegraphics[width=\linewidth]{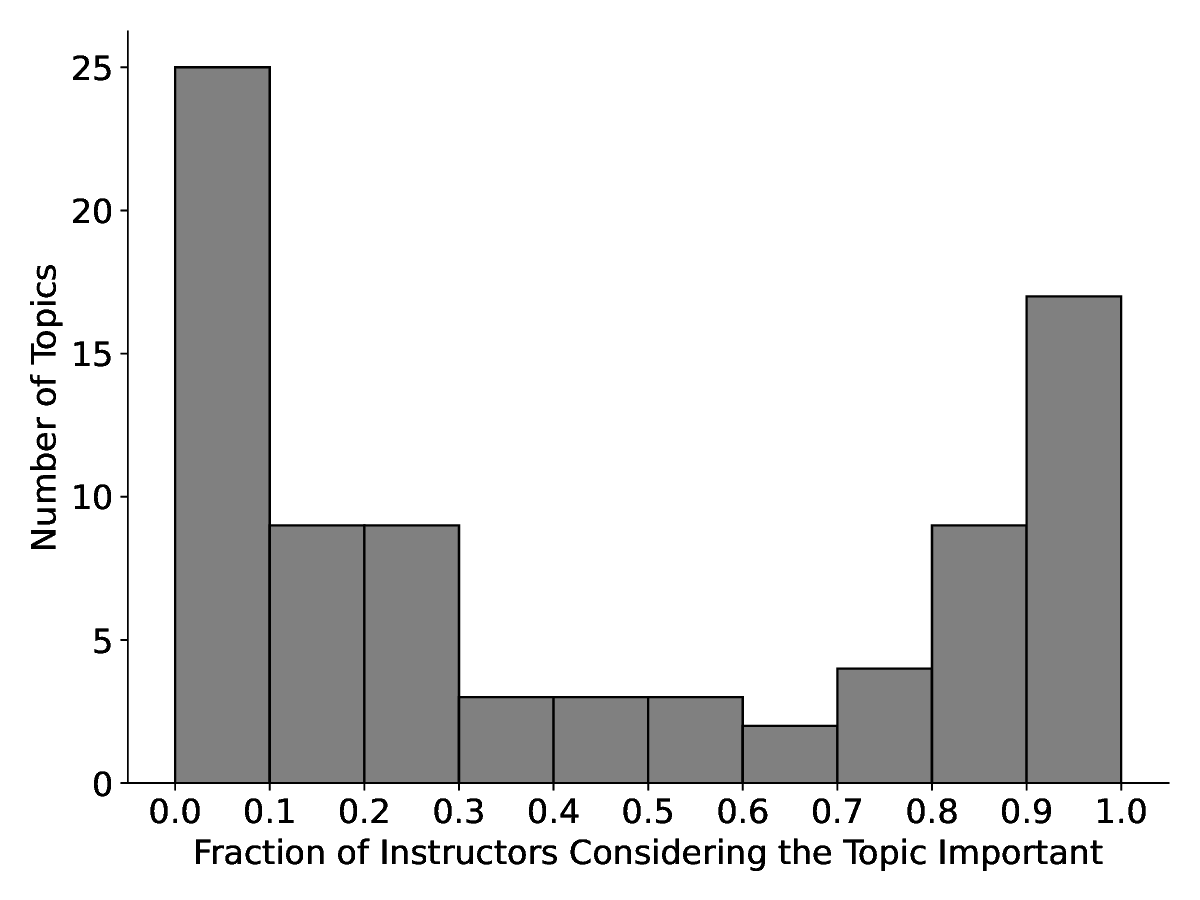}
    \caption{Distribution of the number of topics reported as important by a given fraction of QM1 instructor responses (N = 14). For example, 25 topics were considered important by less than 10\% of instructors.}
    \label{fig:QM1_topic_percs_median}
\end{figure}

\begin{table*}%[b!]
    \centering
    \caption{List of significant topics and percentage of QM1 instructors (N=14) that listed each topic as important. Inclusion was determined by at least 50\% of surveyed instructors considering the topic to be important.}
    \label{tab:QM1_Core}
         \begin{tabular}{p{6 cm} | p{1 cm} | p{6 cm} | p{1 cm}}
          \hline \hline \textbf{Topic} & \textbf{I (\%)} & \textbf{Topic} & \textbf{I  (\%)} \\
         \hline 
            Normalization & 100 & Finite Square Well & 86 \\
Angular Momentum & 100 & Radial Wavefunctions & 86 \\
Hermitian Operators/Observables & 100 & Spherical Harmonics & 86 \\
Time-Indep. Schrödinger Equation & 100 & Eigenstates, Eigenvalues, Eigenequations & 86 \\
Uncertainty Principle & 100 & Linear Algebra Skill & 86 \\
Free Particles & 100 & Probability & 86 \\
Harmonic Oscillator & 100 & Time Evolution & 86 \\
Hydrogen Atom & 100 & Reflection and Transmission & 86 \\
Infinite Square Well & 100 & Matrix Formulation of QM & 79 \\
Expectation Values & 93 & Time-Dep. Schrödinger Equation & 79 \\
Commutators & 93 & Minimum Uncertainty Wavepackets & 71 \\
Momentum & 93 & Stern-Gerlach Experiments & 71 \\
QM Postulates & 93 & Measurement & 64 \\
Raising and Lowering Operators & 93 & Delta-Function Wells and Barriers & 64 \\
Stationary, Bound, Scattering States & 93 & Legendre Polynomials & 57 \\
QM in 3D & 93 & Energy-Time Uncertainty Principle & 57 \\
Spin & 93 & Hermite Polynomials & 50 \\
Hilbert Space & 86 & &\\

         \hline 
    \end{tabular}
\end{table*}

Overall, a total of of 14 respondents reported having taught only a QM1 course. To identify core content areas covered in the QM1 courses in our dataset, we focus on topics considered to be important by instructors. To narrow the focus to topics with a significant amount of consensus, we then needed to select a threshold to delineate topics covered frequently from those covered more sporadically. Different thresholds have been used in prior instructor content surveys (e.g., Refs.~\cite{meyer2024introductory, rainey2020designing}) and all are ultimately somewhat arbitrary. We decide to use a threshold of 50\%, meaning at least half of the instructors in our sample considered the topic to be important. This threshold provides a nice separation between the slightly bimodal distribution of responses shown in Figure \ref{fig:QM1_topic_percs_median}. However, the reader can feel free to enforce a stricter or looser threshold depending on their specific interests (see Appendix~\ref{app:topic}). The topics identified as important by QM1 instructors, based on our 50\% threshold, are listed in Table \ref{tab:QM1_Core}. Together, these topics represent those with a degree of consensus amongst our respondents as appropriate for inclusion in a QM1 course.

To identify additional topics likely to be included in QM2 (and/or QM3 in the case of quarter systems), we turn to the responses from instructors who had taught these courses. There were only four instructors who responded to our survey that taught just a QM2 or QM3 course without having ever taught QM1. However, there were also 18 instructors who responded for both QM1 and QM2. We combined the responses for these group, resulting in a total of 22 instructors with experience teaching more advanced QM topics. Since our goal is not to delineate QM1 topics from QM2 topics but rather to produce a more general sense of topics included in our undergraduate quantum curriculum broadly, we also removed from consideration topics already identified as important for QM1 (Table \ref{tab:QM1_Core}). In this way, we are able to focus on the remaining topics in our list and determine those that are consistently taught above and beyond the QM1 level by focusing on courses likely to include those more more advanced topics. 

For consistency, we use the same 50\% threshold for determining significant topics used for the QM1 topics. Topics that highlight at this level are listed in Table \ref{tab:QM2_Core}. Using the 50\% threshold, we have identified 35 topics in QM1 and 21 additional topics in QM2. This leaves 28 topics from our original list that do not pass the threshold in either group. Of these topics, two of them (Differential Equations and Complex Numbers/Functions) were assumed to be prior knowledge by at least half of the instructors in each group. Additionally, there were only two topics on our list where more than half of QM2 instructors reported not covering them at all: Group Theory and the Hartree-Fock Method. Many of the remaining topics were considered peripheral, rather than important, by a considerable portion of instructors. For example, the Stark Effect was considered important by 36\% of QM2 instructors but peripheral by 50\%. We direct readers interested in topics receiving \emph{any} coverage (i.e., topics considered important or peripheral, rather than just those considered important) to Appendix~\ref{app:topic}. 

\begin{table*}[t!]
    \centering
        \caption{List of significant topics and percentage of QM2 instructors (N=22) that listed each topic as important. Inclusion was determined by at least 50\% of surveyed instructors considering the topic to be important.}
    \label{tab:QM2_Core}
    \begin{tabular}{p{7 cm} | p{1 cm} | p{7cm} | p{1 cm}}
         \hline \hline \textbf{Topic} & \textbf{I (\%)} & \textbf{Topic} & \textbf{I (\%)} \\
         \hline Non-Deg. Time-Indep. Perturbation Theory & 91 & Fine Structure Correction & 59 \\
Deg. Time-Indep. Perturbation Theory & 82 & Zeeman Effect & 59 \\
Identical Particle Systems & 82 & Infinite Spherical Well & 55 \\
Variational Principle & 77 & Schrödinger/Heisenberg/Interaction Picture & 50 \\
Bosons & 77 & Selection Rules & 50 \\
Addition of Angular Momenta & 77 & Symmetries and Conservation Laws & 50 \\
Fermions & 77 & EM Interactions & 50 \\
Electron in a Magnetic Field & 73 & Exchange Interaction & 50 \\
Time-Dep. Perturbation Theory & 68 & Radiation and Emission/Absorption & 50 \\
Unitary Operators & 64 & Entanglement & 50 \\
Transitions & 64 & & \\

         \hline 
    \end{tabular}
\end{table*}

\begin{figure}[b]
    \centering
    \includegraphics[width=\linewidth]{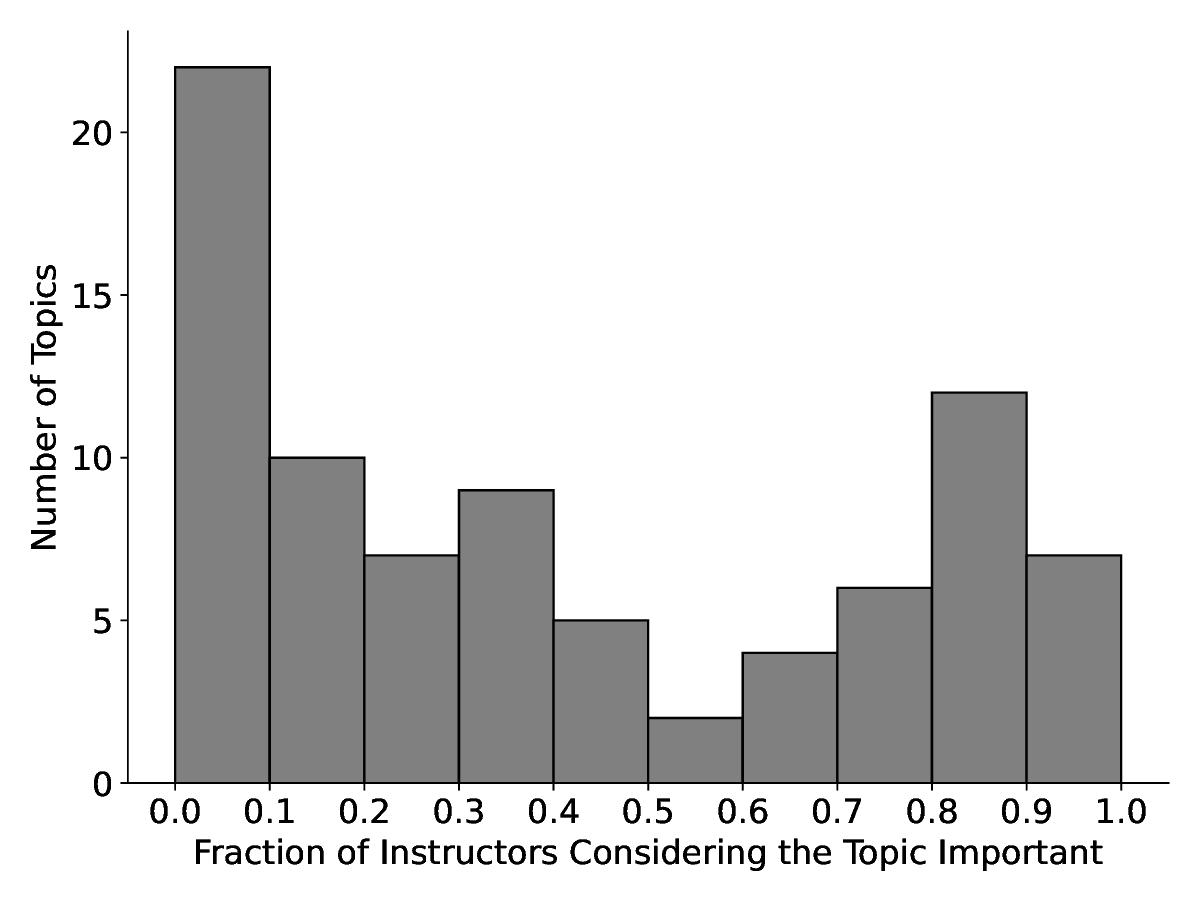}
    \caption{Distribution of the number of topics reported as important by a given fraction of single-semester QM instructor responses (N = 40). For example, 22 topics were considered important by less than 10\% of instructors.}
    \label{fig:QMSS_topic_percs_median}
\end{figure}

Next, we consider topical coverage in single-semester QM courses. We first consider Figure~\ref{fig:QMSS_topic_percs_median}, which shows a histogram of the number of topics binned by the percentage of single-semester QM instructors considering the topic important. In comparison to Figure~\ref{fig:QM1_topic_percs_median} (the equivalent distribution for QM1 instructors), the distribution shows slightly less agreement between instructors. This can be quantified by considering the number of topics where at least 90\% or less than 10\% of instructors consider the topic important. In the sample of QM1 instructors, 42 of the 84 topics were in this range. In the sample of single-semester instructors, 29 of the topics were in this range, showing less agreement among instructors about which topics are important or unimportant. 

We continue by identifying significant topics in single-semester QM courses. Using the same method and threshold as for QM1, we identify the list of topics shown in Table \ref{tab:QMSS_Core}. When comparing the results of QM1 courses to single-semester QM courses, we note that there is considerable overlap in significant topics. The only topic that was considered a significant topic in single-semester QM courses but not in QM1 courses was Addition of Angular Momenta; this was considered important by 43\% of QM1 instructors. There were five topics that were considered significant in QM1 courses but not in single-semester QM courses: Energy-Time Uncertainty Principle (48\%), Legendre Polynomials (43\%), Delta-Function Wells and Barriers (40\%), Hermite Polynomials (32\%), and Minimum Uncertainty Wavepackets (30\%), where the parenthetical indicates the percentage of single-semester QM instructors considering the topic important. Of course, these regions of overlap would change with a different threshold, and the differences could certainly be sample dependent. However, this analysis supports the conclusion that single-semester courses largely align with QM1 courses in terms of topical coverage, albeit with somewhat more variability between instructors. 

\begin{table*}[t!]
    \centering
        \caption{List of significant topics and percentage of single-semester QM instructors (N=40) that listed each topic as important. Inclusion was determined by at least 50\% of surveyed instructors considering the topic to be important.}
    \label{tab:QMSS_Core}
    \begin{tabular}{p{6 cm} | p{1 cm} | p{6cm} | p{1 cm}}
         \hline \hline \textbf{Topic} & \textbf{I (\%)} & \textbf{Topic} & \textbf{I (\%)} \\
         \hline Normalization & 95 & QM Postulates & 80 \\
Commutators & 95 & QM in 3D & 80 \\
Hermitian Operators/Observables & 93 & Momentum & 80 \\
Time-Indep. Schrödinger Equation & 93 & Radial Wavefunctions & 78 \\
Finite Square Well & 93 & Time-Dep. Schrödinger Equation & 73 \\
Harmonic Oscillator & 90 & Probability & 73 \\
Free Particles & 90 & Measurement & 70 \\
Hydrogen Atom & 88 & Time Evolution & 70 \\
Expectation Values & 88 & Raising and Lowering Operators & 70 \\
Infinite Square Well & 88 & Stern-Gerlach Experiments & 65 \\
Spin & 85 & Linear Algebra Skill & 65 \\
Stationary, Bound, Scattering States & 85 & Reflection and Transmission & 60 \\
Uncertainty Principle & 83 & Matrix Formulation of QM & 60 \\
Eigenstates, Eigenvalues, Eigenequations & 82 & Hilbert Space & 58 \\
Spherical Harmonics & 83 & Addition of Angular Momenta & 50 \\
Angular Momentum & 83 & & \\

         \hline 
    \end{tabular}
\end{table*}

\section{Spins vs. Wavefunctions-first approaches}

\begin{figure*}[]
    \centering
    \includegraphics[width=\linewidth]{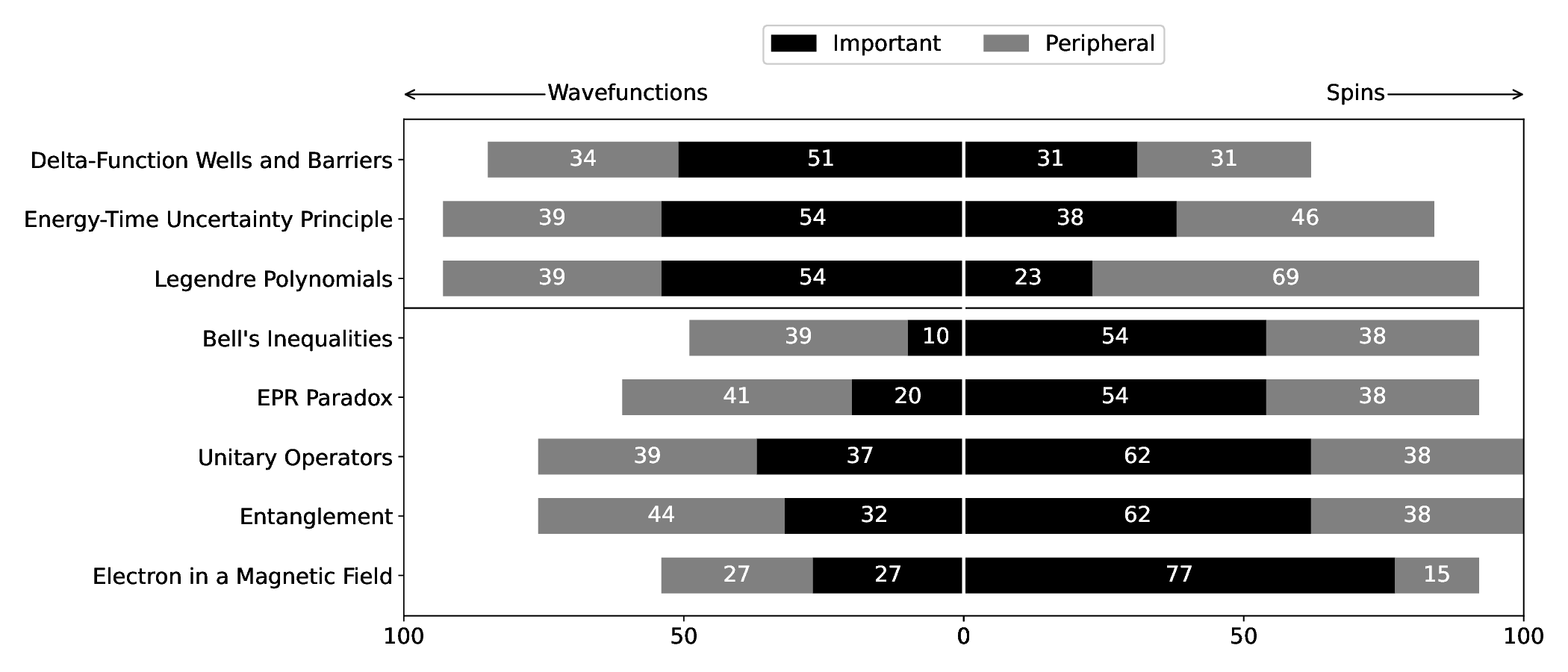}
    \caption{Plot of topics that were significant for either spins-first (N=13) courses or wavefunctions-first courses (N=41) but not both, along with the percentage of instructors reporting each topic as important or peripheral in each group. Wavefunctions-first courses are shown left of center and spins-first courses are shown right of center. Note that QM1 and single semester/quarter QM courses have been combined for this analysis.}
    \label{fig:spins_wfs}
\end{figure*}

While the previous section only addressed differences in topical coverage between QM1, QM2, and single-semester QM, it is also worth considering potential differences between spins-first and wavefunctions-first courses. We consider a course to be spins-first if it only uses a spins-first textbook (McIntyre or Townsend) or the instructor identified the course as spins-first. All other courses are considered to be wavefunctions-first.\footnote{We default to wavefunctions-first because it has historically dominated QM instruction.} Additionally, we focus our attention on QM1 and single-semester/single-quarter QM courses. Based on the results in the previous section, we consider it reasonable to combine those two course categories for this analysis. Thus, the data included in this section includes 13 spins-first courses and 41 wavefunctions-first courses. 

Our analysis follows the procedure described in the previous section. We first construct lists of topics considered important by at least 50\% of instructors in each group. We observe an overlap of 30 topics, all of which also showed up as significant topics for both QM1 and single-semester QM. We additionally observe eight topics that show up as a significant topic in one group but not the other. To further investigate potential differences in coverage for these topics, we also consider the percentage of instructors in each group considering the topic to be peripheral. These topics, along with the percentage of instructors who considered them important or peripheral in each group, are shown in Figure~\ref{fig:spins_wfs}.

Here, it is not our goal to argue that we see statistically significant differences in topical coverage between spins-first and wavefunctions-first courses. Rather, it is interesting to consider the trends we observe in our dataset and how they might inform future analyses. We observe notable overlap between the spins-first and wavefunctions-first courses in our sample: 30 topics surpassed the threshold in both groups, 46 topics did not surpass the threshold in either group, and 8 topics surpassed the threshold in one group but not the other. When considering the eight topics presented in Figure~\ref{fig:spins_wfs}, we note that several of the topics show less discrepancy when considering the percentage of instructors who consider the topic to be either important or peripheral. For example, the energy-time uncertainty principle and Legendre polynomials see similar levels of total coverage despite being considered important by a greater percentage of instructors of wavefunctions-first courses. The electron in a magnetic field, on the other hand, is covered by almost all spins-first instructors and approximately half of wavefunctions-first instructors. Further, 77\% of spins-first instructors in our sample consider it to be an important topic, while only 27\% of wavefunctions-first instructors do. This result is not surprising, as spins-first textbooks typically begin with a discussion of the Stern-Gerlach experiment, which requires a discussion of electrons in magnetic fields. Bell's inequalities shows a similar discrepancy between groups, with 54\% of spins-first instructors considering it important, compared to only 10\% of wavefunctions-first instructors. 

Historically, the comparison of spins-first versus wavefunctions-first approaches has focused on broader conceptual understanding of QM (e.g., \cite{riihiluoma2025sf}), rather than topical coverage. However, instructors wishing to adopt one approach over the other may be concerned with differences in content coverage. Additionally, it is important for assessment developers to have this information when constructing assessments that might be used to compare spins-first and wavefunctions-first approaches.

% CONCLUSION
\section{\label{sec:conclusion}Discussion}

In this paper, we reported the results from a study looking to understand the scope and variability in undergraduate quantum mechanics education across the US. Utilizing a mixed methods approach, we designed a survey and distributed it to QM instructors across the U.S. who taught an upper-level QM course in the last five years, ultimately receiving 76 complete responses. These results highlight several important considerations for the PER community engaged in quantum education research and materials development.

We found that most of the surveyed instructors came from semester-based institutions where a majority of those taught a single-semester QM course rather than one of a multi-course QM sequence. Instructors of stand-alone quantum courses have to make difficult decisions regarding what content and topics they focus on given their more limited timeframe with the inevitable result that they cannot cover all the topics that might be touched on in a multi-semester course. Indeed, we found a significant degree of overlap between what is covered in a single semester course and the first semester of a multi-semester sequence, suggesting that in most cases, a single-semester quantum course is not covering more advanced quantum content. This finding has particularly important implications with respect to graduate education. Particularly at larger R1 schools, it is often implicitly assumed that students enter graduate school with a full undergraduate quantum sequence under their belt. Our results suggest that is often not the case, and graduate instructors should carefully consider what they can assume regarding students' background going into their graduate courses. 

We found that prerequisites for undergraduate QM courses are fairly consistent, with modern physics, differential equations, mathematical methods for physicists, and linear algebra as the top four most commonly listed prerequisites in our dataset. Additionally, Griffiths' text was, by a large margin, the most common textbook, used by roughly 60\% of our respondents. As Griffiths is a wavefunctions-first text, this finding suggests that a wavefunctions-first approach is still the most dominant instructional approach. However, the next two most common texts, McIntyre and Townsend, are spins-first texts utilized by close to a third of our respondents. This suggests that spins-first instruction is seeing use throughout the country. Studies comparing the efficacy of these two approaches to QM instruction are rare \cite{riihiluoma2025sf}, and do not show a clear signal favoring either approach. Ultimately, the value of one approach over the other is likely tied to deeper goals around QM education. For example, the rise of QISE and the trend towards including quantum computation in undergraduate QM may drive some instructors to favor a spins-first approach. However, the tension between these two approaches can present a challenge for both curriculum and assessment developers. While we observe significant overlap in content coverage between the two approaches, our findings also suggest slightly different priorities in terms of content coverage.

Our results also shed interesting light on the teaching methodologies used in undergraduate QM courses. We found that the two most commonly used methodologies were traditional lecture and interactive lecture, which is likely unsurprising. The frequency of these two approach, which were used by a similar fraction of our respondents, however, suggests an interesting trend. As an upper-division course taken primarily by juniors and seniors, conventional wisdom might have lead one to assume traditional lecture would be the dominant form of instruction. However, we observe more instructors using non-traditional instruction (including interactive lecture, a flipped classroom, studio instruction) than traditional instruction. Importantly, a survey methodology will tend to over-sample instructors who have interest in participating in education research and time available to do so. However, our results still suggest that active engagement has been adopted by a large fraction of instructors even in this relatively advanced course. We interpret this as an encouraging finding particularly in light of previous literature suggesting less consistent uptake in research-based instructional strategies \cite{henderson2007barriers}.  

In addition to information on their overall instructional approach, we also investigated the specific  pedagogical resources instructors used in their courses. We found the three most commonly used resources included in-class group work, pre-class readings, and simulations of quantum phenomena to aid in learning. The use of exam and homework corrections was also quite common with roughly a third of respondents reporting each, suggesting that many instructors are incorporating explicit reflective practices within their courses. Concept questions and tutorials were reported at a similar level and many other resources such as demos, group exams, and oral exams were all reported at less often. Overall, our findings suggest that the types of instructional resources used by QM instructors are quite varied and there is potentially demand for new and innovative curricular materials, particularly those that can be inserted into a course without significant modification of the overall course structure. 

We also presented a breakdown of topical content coverage of instructors in our dataset, breaking down the topics list into those reported primarily for QM1 and those for QM2 and beyond. This distinction was not aimed at defining what `should' be taught in each of these course, but rather only to maximize our insight into which topics have a degree of consensus across the full scope of undergraduate QM instruction. This list of topics can serve as a resource for the PER community to help understand the status of QM education, identify gaps in the current resources available to instructors, and guide the development of new curricular and assessment tools.

Finally, we compared content priorities for instructors teaching spins-first and wavefunctions-first QM1 courses or single-semester courses. Our results showed considerable overlap between the two approaches; however, we observed some variance between the two groups. Of the differences we did observe, some were primarily of emphasis (i.e., whether they considered the topic important or merely peripheral), while a few topics saw different levels of overall coverage. Considering potential differences in topical coverage will be important for future analyses comparing spins-first and wavefunctions-first approaches, as well as for QM instructors deciding between the two strategies. 

This work is an important step in understanding the landscape of quantum education in the U.S. at a time when quantum education is more scrutinized than ever. Now is the time to examine what and how we are teaching upper-level undergraduate QM so we can better prepare a quantum workforce capable of supporting the second quantum revolution. 

% ACKNOWLEDGEMENTS
\acknowledgments{} 

We thank Steven Pollock for valuable feedback on the manuscript. This work was supported by funding from the Department of Physics at University of Colorado Boulder, the National Science Foundation DUE Grant No. 2143976, and a National Science Foundation Graduate Research Fellowship.
\appendix
\section{Additional Textbooks}\label{app:details}

\begin{table*}[h!]
\caption{Textbooks not included in Figure \ref{fig:textbooks_all} because they were only selected by one instructor.}
    \centering
    \begin{tabular}{l @{\hspace{15pt}} l}
    \hline \hline
     Textbook & Author \\ \hline
     The Meaning of Quantum Theory & Baggott \\
     The Physics of Quantum Mechanics & Binney and Skinner \\
     Quantum Physics: States Observables and Their Time Evolution & Bohm, Kielanowski, and Mainland \\
     Quantum Mechanics & Cohen-Tannoudji, Diu, and Laloë \\
     Quantum Mechanics & Fitzpatrick \\
     Quantum Mechanics: A Conceptual Approach & Hameka \\
     Quantum Mechanics Non-Relativistic Theory & Landau and Lifshitz \\
     Quantum Mechanics & Merzbacher \\
     Quantum Processes, Systems, and Information & Schumacher and Westmoreland \\
     Invitation to Quantum Mechanics & Styer \\
     The Physics of Quantum Mechanics & Styer \\
     \hline 
    \end{tabular}
    \label{tab:my_label}
\end{table*}

\section{Instructor Survey Content}\label{app:survey}

\begin{figure*}[h]
    \centering
    \includegraphics[width=0.9\linewidth]{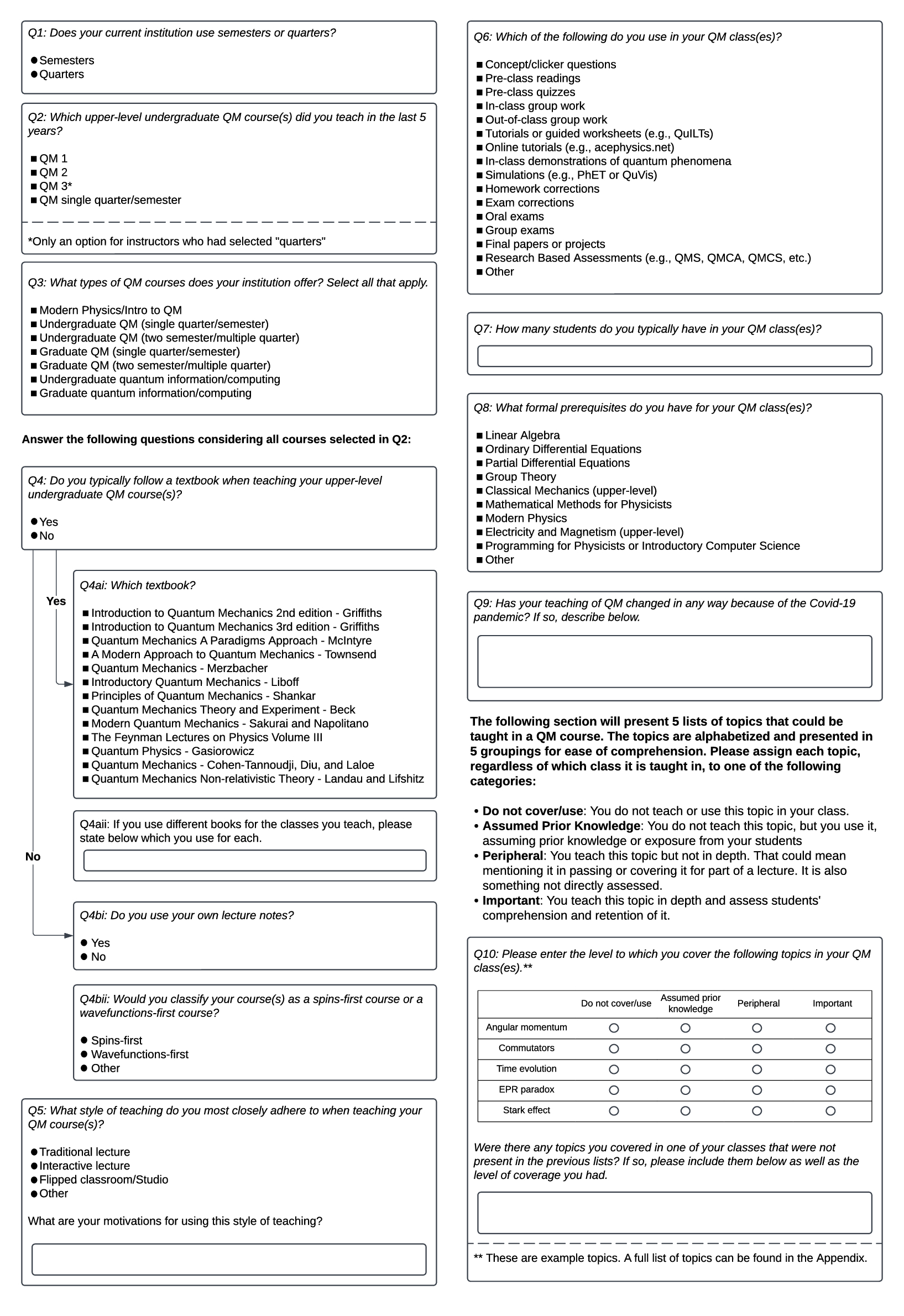}
    \caption{Complete survey, showing survey flow and answer options. The text under the dashed lines in Q2 and Q10 is provided to help with interpretation and was not present in the original survey.}
    \label{fig:enter-label}
\end{figure*}

\section{Topic Coverage Percentages}\label{app:topic}

\begin{table*}[p!]
    \caption{Percentage of QM1 teaching instructors (N=14) who responded for each coverage category. APK stands for Assumed Prior Knowledge, and DNC stands for Do Not Cover. Rows may not add up to 100 due to rounding.}
    \centering
    \begin{tabular}{p{6 cm} | p{1.5 cm} | p{1.5 cm} | p{1 cm} | p{1 cm}}
         \hline \hline \textbf{Topic} & \textbf{Important} & \textbf{Peripheral} & \textbf{APK} & \textbf{DNC} \\
         \hline Normalization & 100 & 0 & 0 & 0 \\
Angular Momentum & 100 & 0 & 0 & 0 \\
Hermitian Operators/Observables & 100 & 0 & 0 & 0 \\
Time-Indep. Schrödinger Equation & 100 & 0 & 0 & 0 \\
Uncertainty Principle & 100 & 0 & 0 & 0 \\
Free Particles & 100 & 0 & 0 & 0 \\
Harmonic Oscillator & 100 & 0 & 0 & 0 \\
Hydrogen Atom & 100 & 0 & 0 & 0 \\
Infinite Square Well & 100 & 0 & 0 & 0 \\
Expectation Values & 93 & 7 & 0 & 0 \\
Commutators & 93 & 7 & 0 & 0 \\
Momentum & 93 & 7 & 0 & 0 \\
QM Postulates & 93 & 7 & 0 & 0 \\
Raising and Lowering Operators & 93 & 7 & 0 & 0 \\
Stationary, Bound, Scattering States & 93 & 7 & 0 & 0 \\
QM in 3D & 93 & 7 & 0 & 0 \\
Spin & 93 & 0 & 0 & 7 \\
Hilbert Space & 86 & 14 & 0 & 0 \\
Finite Square Well & 86 & 14 & 0 & 0 \\
Radial Wavefunctions & 86 & 14 & 0 & 0 \\
Spherical Harmonics & 86 & 14 & 0 & 0 \\
Eigenstates, Eigenvalues, Eigenequations & 86 & 7 & 7 & 0 \\
Linear Algebra Skill & 86 & 7 & 7 & 0 \\
Probability & 86 & 7 & 7 & 0 \\
Time Evolution & 86 & 7 & 0 & 7 \\
Reflection and Transmission & 86 & 0 & 0 & 14 \\
Matrix Formulation of QM & 79 & 14 & 0 & 7 \\
Time-Dep. Schrödinger Equation & 79 & 14 & 0 & 7 \\
Minimum Uncertainty Wavepackets & 71 & 29 & 0 & 0 \\
Stern-Gerlach Experiments & 71 & 14 & 7 & 7 \\
Measurement & 64 & 29 & 7 & 0 \\
Delta-Function Wells and Barriers & 64 & 21 & 0 & 14 \\
Legendre Polynomials & 57 & 36 & 0 & 7 \\
Energy-Time Uncertainty Principle & 57 & 36 & 0 & 7 \\
Hermite Polynomials & 50 & 29 & 0 & 21 \\
Infinite Spherical Well & 43 & 50 & 0 & 7 \\
Electron in a Magnetic Field & 43 & 21 & 0 & 36 \\
Addition of Angular Momenta & 43 & 14 & 0 & 43 \\
Special Functions & 36 & 50 & 0 & 14 \\
Unitary Operators & 36 & 43 & 7 & 14 \\
Symmetries and Conservation Laws & 36 & 14 & 0 & 50 \\
Periodic Table of Elements & 29 & 36 & 0 & 36 \\
Bosons & 29 & 0 & 7 & 64 \\
Deg. Time-Indep. Perturbation Theory & 21 & 21 & 0 & 57 \\
Coherent States & 21 & 14 & 7 & 57 \\
Identical Particle Systems & 21 & 14 & 0 & 64 \\
Frobenius Method & 21 & 7 & 0 & 71 \\
Fermions & 21 & 0 & 7 & 71 \\
Non-Deg. Time-Indep. Perturbation Theory & 21 & 0 & 0 & 79 \\
Born Approximation & 21 & 0 & 0 & 79 \\
Entanglement & 14 & 57 & 0 & 29 \\
Clebsch-Gordan Coefficients & 14 & 43 & 0 & 43 \\
Scattering & 14 & 29 & 0 & 57 \\
Density Operators/Mixed states & 14 & 21 & 0 & 64 \\
Periodic Potentials & 14 & 21 & 0 & 64 \\
Zeeman Effect & 14 & 14 & 7 & 64 \\
EM Interactions & 14 & 14 & 0 & 71 \\
Differential Equations & 14 & 7 & 79 & 0 \\
         \hline
    \end{tabular}
   
    \label{tab:QM1_Topic_Percs_1}
\end{table*}

\begin{table*}[t!]
    \caption{Percentage of QM1 teaching instructors (N=14) who responded for each coverage category. APK stands for Assumed Prior Knowledge, and DNC stands for Do Not Cover. Rows may not add up to 100 due to rounding.}
    \centering
    \begin{tabular}{p{6 cm} | p{1.5 cm} | p{1.5 cm} | p{1 cm} | p{1 cm}}
         \hline \hline \textbf{Topic} & \textbf{Important} & \textbf{Peripheral} & \textbf{APK} & \textbf{DNC} \\
         \hline Complex Numbers/Functions & 14 & 7 & 71 & 7 \\
EPR Paradox & 7 & 64 & 0 & 29 \\
Bell's Inequalities & 7 & 57 & 0 & 36 \\
Many-Particle Systems & 7 & 29 & 0 & 64 \\
Schrödinger/Heisenberg/Interaction Picture & 7 & 21 & 0 & 71 \\
Stark Effect & 7 & 21 & 0 & 71 \\
Transitions & 7 & 21 & 0 & 71 \\
Exchange Interaction & 7 & 14 & 0 & 79 \\
WKB Approximation & 7 & 7 & 7 & 79 \\
Fine Structure Correction & 7 & 7 & 0 & 86 \\
Hyperfine Structure Correction & 7 & 7 & 0 & 86 \\
Time-Dep. Perturbation Theory & 7 & 7 & 0 & 86 \\
Hund's Rule & 7 & 7 & 0 & 86 \\
Quantum Information and Computation & 7 & 7 & 0 & 86 \\
Bloch Sphere & 0 & 43 & 0 & 57 \\
Selection Rules & 0 & 29 & 0 & 71 \\
Variational Principle & 0 & 21 & 7 & 71 \\
Group Theory & 0 & 21 & 0 & 79 \\
Fermi's Golden Rule & 0 & 21 & 0 & 79 \\
Molecules/Molecular Physics & 0 & 21 & 0 & 79 \\
Einstein Coefficients & 0 & 7 & 7 & 86 \\
Hartree-Fock Method & 0 & 7 & 0 & 93 \\
Partial Wave Decomposition & 0 & 7 & 0 & 93 \\
Radiation and Emission/Absorption & 0 & 7 & 0 & 93 \\
Wigner-Eckart Theorem & 0 & 0 & 7 & 93 \\
Solids & 0 & 0 & 0 & 100 \\
         \hline
    \end{tabular}
    \label{tab:QM1_Topic_Percs_2}
\end{table*}

\begin{table*}[b!]
    \caption{Percentage of QM2 teaching instructors (N=22) who responded to each coverage category. APK stands for Assumed Prior Knowledge, and DNC stands for Do Not Cover. Note that 18/22 instructors in this sample were responding for both QM1 and QM2. Rows may not add up to 100 due to rounding.}
    \centering
    \begin{tabular}{p{6 cm} | p{1.5 cm} | p{1.5 cm} | p{1 cm} | p{1 cm}}
         \hline \hline \textbf{Topic} & \textbf{Important} & \textbf{Peripheral} & \textbf{APK} & \textbf{DNC} \\
         \hline Raising and Lowering Operators & 100 & 0 & 0 & 0 \\
Harmonic Oscillator & 100 & 0 & 0 & 0 \\
Angular Momentum & 95 & 5 & 0 & 0 \\
Commutators & 95 & 0 & 5 & 0 \\
Hydrogen Atom & 95 & 0 & 5 & 0 \\
Radial Wavefunctions & 91 & 9 & 0 & 0 \\
Stationary, Bound, Scattering States & 91 & 5 & 5 & 0 \\
Non-Deg. Time-Indep. Perturbation Theory & 91 & 5 & 0 & 5 \\
Hermitian Operators/Observables & 91 & 0 & 9 & 0 \\
Time-Indep. Schrödinger Equation & 91 & 0 & 9 & 0 \\
Uncertainty Principle & 91 & 0 & 9 & 0 \\
Spin & 91 & 0 & 9 & 0 \\
QM in 3D & 86 & 14 & 0 & 0 \\
Time Evolution & 86 & 9 & 5 & 0 \\
Spherical Harmonics & 86 & 9 & 5 & 0 \\
Hilbert Space & 86 & 5 & 9 & 0 \\
Matrix Formulation of QM & 86 & 5 & 9 & 0 \\
Momentum & 86 & 5 & 9 & 0 \\
Free Particles & 86 & 5 & 9 & 0 \\
QM Postulates & 86 & 0 & 14 & 0 \\
Expectation Values & 86 & 0 & 9 & 5 \\
Measurement & 82 & 14 & 5 & 0 \\
Time-Dep. Schrödinger Equation & 82 & 14 & 5 & 0 \\
Deg. Time-Indep. Perturbation Theory & 82 & 14 & 0 & 5 \\
Identical Particle Systems & 82 & 14 & 0 & 5 \\
        \hline
    \end{tabular}
    \label{tab:QM2_Topic_Percs_1}
\end{table*}

\begin{table*}[p]
    \caption{Percentage of QM2 teaching instructors (N=22) who responded to each coverage category. APK stands for Assumed Prior Knowledge, and DNC stands for Do Not Cover. Note that 18/22 instructors in this sample were responding for both QM1 and QM2. Rows may not add up to 100 due to rounding.}
    \centering
    \begin{tabular}{p{6 cm} | p{1.5 cm} | p{1.5 cm} | p{1 cm} | p{1 cm}}
         \hline \hline \textbf{Topic} & \textbf{Important} & \textbf{Peripheral} & \textbf{APK} & \textbf{DNC} \\
         \hline Finite Square Well & 82 & 9 & 5 & 5 \\
Fermions & 77 & 18 & 0 & 5 \\
Addition of Angular Momenta & 77 & 14 & 5 & 5 \\
Bosons & 77 & 14 & 0 & 9 \\
Infinite Square Well & 77 & 9 & 9 & 5 \\
Eigenstates, Eigenvalues, Eigenequations & 77 & 5 & 18 & 0 \\
Normalization & 77 & 5 & 18 & 0 \\
Variational Principle & 77 & 5 & 0 & 18 \\
Electron in a Magnetic Field & 73 & 18 & 5 & 5 \\
Probability & 73 & 9 & 18 & 0 \\
Stern-Gerlach Experiments & 68 & 23 & 9 & 0 \\
Time-Dep. Perturbation Theory & 68 & 18 & 0 & 14 \\
Linear Algebra Skill & 68 & 5 & 27 & 0 \\
Hermite Polynomials & 64 & 32 & 0 & 5 \\
Legendre Polynomials & 64 & 32 & 0 & 5 \\
Unitary Operators & 64 & 27 & 9 & 0 \\
Transitions & 64 & 23 & 0 & 14 \\
Delta-Function Wells and Barriers & 64 & 18 & 9 & 9 \\
Fine Structure Correction & 59 & 27 & 0 & 14 \\
Zeeman Effect & 59 & 23 & 0 & 18 \\
Reflection and Transmission & 55 & 32 & 9 & 5 \\
Infinite Spherical Well & 55 & 18 & 0 & 27 \\
Selection Rules & 50 & 45 & 0 & 5 \\
Entanglement & 50 & 41 & 5 & 5 \\
Minimum Uncertainty Wavepackets & 50 & 41 & 5 & 5 \\
Symmetries and Conservation Laws & 50 & 41 & 0 & 9 \\
EM Interactions & 50 & 41 & 0 & 9 \\
Radiation and Emission/Absorption & 50 & 41 & 0 & 9 \\
Energy-Time Uncertainty Principle & 50 & 36 & 5 & 9 \\
Schrödinger/Heisenberg/Interaction Picture & 50 & 36 & 0 & 14 \\
Exchange Interaction & 50 & 36 & 0 & 14 \\
Many-Particle Systems & 45 & 45 & 0 & 9 \\
Clebsch-Gordan Coefficients & 45 & 41 & 5 & 9 \\
Fermi's Golden Rule & 45 & 32 & 0 & 23 \\
Hyperfine Structure Correction & 45 & 32 & 0 & 23 \\
WKB Approximation & 45 & 32 & 0 & 23 \\
Scattering & 45 & 32 & 0 & 23 \\
Special Functions & 41 & 45 & 9 & 5 \\
Stark Effect & 36 & 50 & 0 & 14 \\
Density Operators/Mixed states & 36 & 45 & 0 & 18 \\
Coherent States & 32 & 59 & 0 & 9 \\
EPR Paradox & 32 & 55 & 5 & 9 \\
Einstein Coefficients & 32 & 45 & 0 & 23 \\
Periodic Potentials & 32 & 32 & 5 & 32 \\
Born Approximation & 32 & 32 & 0 & 36 \\
Differential Equations & 27 & 23 & 50 & 0 \\
Complex Numbers/Functions & 27 & 5 & 64 & 5 \\
Bell's Inequalities & 23 & 55 & 5 & 18 \\
Periodic Table of Elements & 23 & 55 & 0 & 23 \\
Solids & 23 & 32 & 5 & 41 \\
Molecules/Molecular Physics & 18 & 50 & 0 & 32 \\
Frobenius Method & 18 & 41 & 0 & 41 \\
Wigner-Eckart Theorem & 18 & 41 & 0 & 41 \\
Quantum Information and Computation & 18 & 36 & 5 & 41 \\
Partial Wave Decomposition & 18 & 36 & 0 & 45 \\
Hund's Rule & 9 & 45 & 0 & 45 \\
Hartree-Fock Method & 9 & 41 & 0 & 50 \\
Group Theory & 9 & 32 & 0 & 59 \\
Bloch Sphere & 5 & 64 & 0 & 32 \\
         \hline
    \end{tabular}
    \label{tab:QM2_Topic_Percs_2}
\end{table*}

\begin{table*}[p!]
    \caption{Percentage of single-semester QM instructors (N=40) who responded for each coverage category. APK stands for Assumed Prior Knowledge, and DNC stands for Do Not Cover. Rows may not add up to 100 due to rounding.}
    \centering
    \begin{tabular}{p{6 cm} | p{1.5 cm} | p{1.5 cm} | p{1 cm} | p{1 cm}}
         \hline \hline \textbf{Topic} & \textbf{Important} & \textbf{Peripheral} & \textbf{APK} & \textbf{DNC} \\
         \hline Normalization & 95 & 5 & 0 & 0 \\
Normalization & 95 & 5 & 0 & 0 \\
Commutators & 95 & 5 & 0 & 0 \\
Hermitian Operators/Observables & 93 & 5 & 0 & 3 \\
Time-Indep. Schrödinger Equation & 93 & 3 & 5 & 0 \\
Finite Square Well & 93 & 3 & 5 & 0 \\
Harmonic Oscillator & 90 & 5 & 3 & 3 \\
Free Particles & 90 & 3 & 5 & 3 \\
Hydrogen Atom & 88 & 10 & 3 & 0 \\
Expectation Values & 88 & 8 & 5 & 0 \\
Infinite Square Well & 88 & 5 & 5 & 3 \\
Spin & 85 & 10 & 5 & 0 \\
Stationary, Bound, Scattering States & 85 & 8 & 5 & 3 \\
Uncertainty Principle & 83 & 18 & 0 & 0 \\
Eigenstates, Eigenvalues, Eigenequations & 83 & 13 & 5 & 0 \\
Spherical Harmonics & 83 & 13 & 3 & 3 \\
Angular Momentum & 83 & 8 & 10 & 0 \\
QM Postulates & 80 & 18 & 0 & 3 \\
QM in 3D & 80 & 13 & 3 & 5 \\
Momentum & 80 & 10 & 8 & 3 \\
Radial Wavefunctions & 78 & 18 & 3 & 3 \\
Time-Dep. Schrödinger Equation & 73 & 25 & 0 & 3 \\
Probability & 73 & 20 & 8 & 0 \\
Measurement & 70 & 28 & 3 & 0 \\
Time Evolution & 70 & 23 & 0 & 8 \\
Raising and Lowering Operators & 70 & 20 & 0 & 10 \\
Stern-Gerlach Experiments & 65 & 23 & 5 & 8 \\
Linear Algebra Skill & 65 & 18 & 15 & 3 \\
Reflection and Transmission & 60 & 23 & 8 & 10 \\
Matrix Formulation of QM & 60 & 20 & 3 & 18 \\
Hilbert Space & 58 & 30 & 3 & 10 \\
Addition of Angular Momenta & 50 & 23 & 10 & 18 \\
Energy-Time Uncertainty Principle & 48 & 43 & 0 & 10 \\
Entanglement & 48 & 38 & 0 & 15 \\
Unitary Operators & 45 & 38 & 0 & 18 \\
Legendre Polynomials & 43 & 50 & 0 & 8 \\
Delta-Function Wells and Barriers & 40 & 38 & 3 & 20 \\
Electron in a Magnetic Field & 38 & 25 & 5 & 33 \\
EPR Paradox & 35 & 33 & 0 & 33 \\
Hermite Polynomials & 33 & 53 & 0 & 15 \\
Non-Deg. Time-Indep. Perturbation Theory & 33 & 5 & 0 & 63 \\
Minimum Uncertainty Wavepackets & 30 & 58 & 0 & 13 \\
Special Functions & 30 & 55 & 0 & 15 \\
Fermions & 30 & 30 & 5 & 35 \\
Bosons & 30 & 28 & 5 & 38 \\
Complex Numbers/Functions & 30 & 20 & 50 & 0 \\
Infinite Spherical Well & 28 & 43 & 0 & 30 \\
Identical Particle Systems & 28 & 23 & 3 & 48 \\
Density Operators/Mixed states & 25 & 35 & 0 & 40 \\
Symmetries and Conservation Laws & 25 & 33 & 3 & 40 \\
Bell's Inequalities & 25 & 33 & 0 & 43 \\
Differential Equations & 25 & 20 & 55 & 0 \\
Deg. Time-Indep. Perturbation Theory & 23 & 13 & 0 & 65 \\
Coherent States & 18 & 33 & 0 & 50 \\
Clebsch-Gordan Coefficients & 18 & 30 & 0 & 53 \\
Hyperfine Structure Correction & 18 & 13 & 0 & 70 \\
Schrödinger/Heisenberg/Interaction Picture & 15 & 33 & 0 & 53 \\
Scattering & 15 & 20 & 3 & 63 \\
Variational Principle & 15 & 13 & 0 & 73 \\
         \hline
    \end{tabular}
   
    \label{tab:QM1_Topic_Percs_1}
\end{table*}

\begin{table*}[t!]
    \caption{Percentage of single-semester QM instructors (N=40) who responded for each coverage category. APK stands for Assumed Prior Knowledge, and DNC stands for Do Not Cover. Rows may not add up to 100 due to rounding.}
    \centering
    \begin{tabular}{p{6 cm} | p{1.5 cm} | p{1.5 cm} | p{1 cm} | p{1 cm}}
         \hline \hline \textbf{Topic} & \textbf{Important} & \textbf{Peripheral} & \textbf{APK} & \textbf{DNC} \\
         \hline Many-Particle Systems & 13 & 28 & 0 & 60 \\
Fine Structure Correction & 13 & 15 & 0 & 73 \\
Quantum Information and Computation & 10 & 35 & 0 & 55 \\
Zeeman Effect & 10 & 28 & 3 & 60 \\
Frobenius Method & 8 & 33 & 0 & 60 \\
Transitions & 8 & 13 & 3 & 78 \\
Bloch Sphere & 5 & 20 & 0 & 75 \\
Born Approximation & 5 & 18 & 3 & 75 \\
Group Theory & 5 & 18 & 0 & 78 \\
Time-Dep. Perturbation Theory & 5 & 3 & 0 & 93 \\
Selection Rules & 3 & 33 & 3 & 63 \\
Periodic Table of Elements & 3 & 30 & 13 & 55 \\
Molecules/Molecular Physics & 3 & 23 & 0 & 75 \\
Exchange Interaction & 3 & 20 & 0 & 78 \\
Radiation and Emission/Absorption & 3 & 18 & 3 & 78 \\
Stark Effect & 3 & 13 & 0 & 85 \\
WKB Approximation & 3 & 13 & 0 & 85 \\
Periodic Potentials & 0 & 23 & 3 & 75 \\
EM Interactions & 0 & 20 & 3 & 78 \\
Fermi's Golden Rule & 0 & 15 & 0 & 85 \\
Solids & 0 & 15 & 0 & 85 \\
Hund's Rule & 0 & 8 & 0 & 93 \\
Wigner-Eckart Theorem & 0 & 5 & 0 & 95 \\
Einstein Coefficients & 0 & 5 & 0 & 95 \\
Hartree-Fock Method & 0 & 3 & 0 & 98 \\
Partial Wave Decomposition & 0 & 0 & 0 & 100 \\

         \hline
    \end{tabular}
    \label{tab:QM1_Topic_Percs_2}
\end{table*}

\newpage

% REFERENCES
\clearpage
\bibliography{references-mod.bib}

\begin{thebibliography}{43}%
\makeatletter
\providecommand \@ifxundefined [1]{%
 \@ifx{#1\undefined}
}%
\providecommand \@ifnum [1]{%
 \ifnum #1\expandafter \@firstoftwo
 \else \expandafter \@secondoftwo
 \fi
}%
\providecommand \@ifx [1]{%
 \ifx #1\expandafter \@firstoftwo
 \else \expandafter \@secondoftwo
 \fi
}%
\providecommand \natexlab [1]{#1}%
\providecommand \enquote  [1]{``#1''}%
\providecommand \bibnamefont  [1]{#1}%
\providecommand \bibfnamefont [1]{#1}%
\providecommand \citenamefont [1]{#1}%
\providecommand \href@noop [0]{\@secondoftwo}%
\providecommand \href [0]{\begingroup \@sanitize@url \@href}%
\providecommand \@href[1]{\@@startlink{#1}\@@href}%
\providecommand \@@href[1]{\endgroup#1\@@endlink}%
\providecommand \@sanitize@url [0]{\catcode `\\12\catcode `\$12\catcode `\&12\catcode `\#12\catcode `\^12\catcode `\_12\catcode `\%12\relax}%
\providecommand \@@startlink[1]{}%
\providecommand \@@endlink[0]{}%
\providecommand \url  [0]{\begingroup\@sanitize@url \@url }%
\providecommand \@url [1]{\endgroup\@href {#1}{\urlprefix }}%
\providecommand \urlprefix  [0]{URL }%
\providecommand \Eprint [0]{\href }%
\providecommand \doibase [0]{https://doi.org/}%
\providecommand \selectlanguage [0]{\@gobble}%
\providecommand \bibinfo  [0]{\@secondoftwo}%
\providecommand \bibfield  [0]{\@secondoftwo}%
\providecommand \translation [1]{[#1]}%
\providecommand \BibitemOpen [0]{}%
\providecommand \bibitemStop [0]{}%
\providecommand \bibitemNoStop [0]{.\EOS\space}%
\providecommand \EOS [0]{\spacefactor3000\relax}%
\providecommand \BibitemShut  [1]{\csname bibitem#1\endcsname}%
\let\auto@bib@innerbib\@empty
%</preamble>
\bibitem [{\citenamefont {Dowling}\ and\ \citenamefont {Milburn}(2003)}]{dowling2003quantum}%
  \BibitemOpen
  \bibfield  {author} {\bibinfo {author} {\bibfnamefont {J.~P.}\ \bibnamefont {Dowling}}\ and\ \bibinfo {author} {\bibfnamefont {G.~J.}\ \bibnamefont {Milburn}},\ }\bibfield  {title} {\bibinfo {title} {Quantum technology: the second quantum revolution},\ }\href@noop {} {\bibfield  {journal} {\bibinfo  {journal} {Philosophical Transactions of the Royal Society of London. Series A: Mathematical, Physical and Engineering Sciences}\ }\textbf {\bibinfo {volume} {361}},\ \bibinfo {pages} {1655} (\bibinfo {year} {2003})}\BibitemShut {NoStop}%
\bibitem [{\citenamefont {Hughes}\ \emph {et~al.}(2022)\citenamefont {Hughes}, \citenamefont {Finke}, \citenamefont {German}, \citenamefont {Merzbacher}, \citenamefont {Vora},\ and\ \citenamefont {Lewandowski}}]{hughes2022assessing}%
  \BibitemOpen
  \bibfield  {author} {\bibinfo {author} {\bibfnamefont {C.}~\bibnamefont {Hughes}}, \bibinfo {author} {\bibfnamefont {D.}~\bibnamefont {Finke}}, \bibinfo {author} {\bibfnamefont {D.-A.}\ \bibnamefont {German}}, \bibinfo {author} {\bibfnamefont {C.}~\bibnamefont {Merzbacher}}, \bibinfo {author} {\bibfnamefont {P.~M.}\ \bibnamefont {Vora}},\ and\ \bibinfo {author} {\bibfnamefont {H.}~\bibnamefont {Lewandowski}},\ }\bibfield  {title} {\bibinfo {title} {Assessing the needs of the quantum industry},\ }\href@noop {} {\bibfield  {journal} {\bibinfo  {journal} {IEEE Transactions on Education}\ }\textbf {\bibinfo {volume} {65}},\ \bibinfo {pages} {592} (\bibinfo {year} {2022})}\BibitemShut {NoStop}%
\bibitem [{\citenamefont {Fox}\ \emph {et~al.}(2020)\citenamefont {Fox}, \citenamefont {Zwickl},\ and\ \citenamefont {Lewandowski}}]{fox2020preparing}%
  \BibitemOpen
  \bibfield  {author} {\bibinfo {author} {\bibfnamefont {M.~F.}\ \bibnamefont {Fox}}, \bibinfo {author} {\bibfnamefont {B.~M.}\ \bibnamefont {Zwickl}},\ and\ \bibinfo {author} {\bibfnamefont {H.~J.}\ \bibnamefont {Lewandowski}},\ }\bibfield  {title} {\bibinfo {title} {Preparing for the quantum revolution: What is the role of higher education?},\ }\href@noop {} {\bibfield  {journal} {\bibinfo  {journal} {Physical Review Physics Education Research}\ }\textbf {\bibinfo {volume} {16}},\ \bibinfo {pages} {020131} (\bibinfo {year} {2020})}\BibitemShut {NoStop}%
\bibitem [{\citenamefont {Dubson}\ \emph {et~al.}(2009)\citenamefont {Dubson}, \citenamefont {Goldhaber}, \citenamefont {Pollock},\ and\ \citenamefont {Perkins}}]{dubson2009faculty}%
  \BibitemOpen
  \bibfield  {author} {\bibinfo {author} {\bibfnamefont {M.}~\bibnamefont {Dubson}}, \bibinfo {author} {\bibfnamefont {S.}~\bibnamefont {Goldhaber}}, \bibinfo {author} {\bibfnamefont {S.}~\bibnamefont {Pollock}},\ and\ \bibinfo {author} {\bibfnamefont {K.}~\bibnamefont {Perkins}},\ }\bibfield  {title} {\bibinfo {title} {Faculty disagreement about the teaching of quantum mechanics},\ }in\ \href@noop {} {\emph {\bibinfo {booktitle} {AIP Conference Proceedings}}},\ Vol.\ \bibinfo {volume} {1179}\ (\bibinfo {organization} {American Institute of Physics},\ \bibinfo {year} {2009})\ pp.\ \bibinfo {pages} {137--140}\BibitemShut {NoStop}%
\bibitem [{\citenamefont {Siddiqui}\ and\ \citenamefont {Singh}(2017)}]{siddiqui2017diverse}%
  \BibitemOpen
  \bibfield  {author} {\bibinfo {author} {\bibfnamefont {S.}~\bibnamefont {Siddiqui}}\ and\ \bibinfo {author} {\bibfnamefont {C.}~\bibnamefont {Singh}},\ }\bibfield  {title} {\bibinfo {title} {How diverse are physics instructors’ attitudes and approaches to teaching undergraduate level quantum mechanics?},\ }\href@noop {} {\bibfield  {journal} {\bibinfo  {journal} {European Journal of Physics}\ }\textbf {\bibinfo {volume} {38}},\ \bibinfo {pages} {035703} (\bibinfo {year} {2017})}\BibitemShut {NoStop}%
\bibitem [{\citenamefont {Griffiths}\ and\ \citenamefont {Schroeter}(2018)}]{griffiths2018introduction}%
  \BibitemOpen
  \bibfield  {author} {\bibinfo {author} {\bibfnamefont {D.~J.}\ \bibnamefont {Griffiths}}\ and\ \bibinfo {author} {\bibfnamefont {D.~F.}\ \bibnamefont {Schroeter}},\ }\href@noop {} {\emph {\bibinfo {title} {Introduction to quantum mechanics}}}\ (\bibinfo  {publisher} {Cambridge University Press},\ \bibinfo {year} {2018})\BibitemShut {NoStop}%
\bibitem [{\citenamefont {McIntyre}(2022)}]{mcintyre2022quantum}%
  \BibitemOpen
  \bibfield  {author} {\bibinfo {author} {\bibfnamefont {D.~H.}\ \bibnamefont {McIntyre}},\ }\href@noop {} {\emph {\bibinfo {title} {Quantum mechanics}}}\ (\bibinfo  {publisher} {Cambridge University Press},\ \bibinfo {year} {2022})\BibitemShut {NoStop}%
\bibitem [{\citenamefont {Townsend}(2000)}]{townsend2000modern}%
  \BibitemOpen
  \bibfield  {author} {\bibinfo {author} {\bibfnamefont {J.~S.}\ \bibnamefont {Townsend}},\ }\href@noop {} {\emph {\bibinfo {title} {A modern approach to quantum mechanics}}}\ (\bibinfo  {publisher} {University Science Books},\ \bibinfo {year} {2000})\BibitemShut {NoStop}%
\bibitem [{\citenamefont {Shankar}(2012)}]{shankar2012principles}%
  \BibitemOpen
  \bibfield  {author} {\bibinfo {author} {\bibfnamefont {R.}~\bibnamefont {Shankar}},\ }\href@noop {} {\emph {\bibinfo {title} {Principles of quantum mechanics}}}\ (\bibinfo  {publisher} {Springer Science \& Business Media},\ \bibinfo {year} {2012})\BibitemShut {NoStop}%
\bibitem [{\citenamefont {Beck}(2012)}]{beck2012quantum}%
  \BibitemOpen
  \bibfield  {author} {\bibinfo {author} {\bibfnamefont {M.}~\bibnamefont {Beck}},\ }\href@noop {} {\emph {\bibinfo {title} {Quantum mechanics: theory and experiment}}}\ (\bibinfo  {publisher} {Oxford University Press},\ \bibinfo {year} {2012})\BibitemShut {NoStop}%
\bibitem [{\citenamefont {McKagan}\ \emph {et~al.}(2020)\citenamefont {McKagan}, \citenamefont {Strubbe}, \citenamefont {Barbato}, \citenamefont {Mason}, \citenamefont {Madsen},\ and\ \citenamefont {Sayre}}]{mckagan2020physport}%
  \BibitemOpen
  \bibfield  {author} {\bibinfo {author} {\bibfnamefont {S.~B.}\ \bibnamefont {McKagan}}, \bibinfo {author} {\bibfnamefont {L.~E.}\ \bibnamefont {Strubbe}}, \bibinfo {author} {\bibfnamefont {L.~J.}\ \bibnamefont {Barbato}}, \bibinfo {author} {\bibfnamefont {B.~A.}\ \bibnamefont {Mason}}, \bibinfo {author} {\bibfnamefont {A.~M.}\ \bibnamefont {Madsen}},\ and\ \bibinfo {author} {\bibfnamefont {E.~C.}\ \bibnamefont {Sayre}},\ }\bibfield  {title} {\bibinfo {title} {Physport use and growth: Supporting physics teaching with research-based resources since 2011},\ }\href@noop {} {\bibfield  {journal} {\bibinfo  {journal} {The Physics Teacher}\ }\textbf {\bibinfo {volume} {58}},\ \bibinfo {pages} {465} (\bibinfo {year} {2020})}\BibitemShut {NoStop}%
\bibitem [{\citenamefont {Cataloglu}\ and\ \citenamefont {Robinett}(2002)}]{cataloglu2002testing}%
  \BibitemOpen
  \bibfield  {author} {\bibinfo {author} {\bibfnamefont {E.}~\bibnamefont {Cataloglu}}\ and\ \bibinfo {author} {\bibfnamefont {R.}~\bibnamefont {Robinett}},\ }\bibfield  {title} {\bibinfo {title} {Testing the development of student conceptual and visualization understanding in quantum mechanics through the undergraduate career},\ }\href@noop {} {\bibfield  {journal} {\bibinfo  {journal} {American Journal of Physics}\ }\textbf {\bibinfo {volume} {70}},\ \bibinfo {pages} {238} (\bibinfo {year} {2002})}\BibitemShut {NoStop}%
\bibitem [{\citenamefont {Sadaghiani}\ and\ \citenamefont {Pollock}(2015)}]{sadaghiani2015quantum}%
  \BibitemOpen
  \bibfield  {author} {\bibinfo {author} {\bibfnamefont {H.~R.}\ \bibnamefont {Sadaghiani}}\ and\ \bibinfo {author} {\bibfnamefont {S.~J.}\ \bibnamefont {Pollock}},\ }\bibfield  {title} {\bibinfo {title} {Quantum mechanics concept assessment: Development and validation study},\ }\href@noop {} {\bibfield  {journal} {\bibinfo  {journal} {Physical Review Special Topics-Physics Education Research}\ }\textbf {\bibinfo {volume} {11}},\ \bibinfo {pages} {010110} (\bibinfo {year} {2015})}\BibitemShut {NoStop}%
\bibitem [{\citenamefont {Marshman}\ and\ \citenamefont {Singh}(2019)}]{marshman2019validation}%
  \BibitemOpen
  \bibfield  {author} {\bibinfo {author} {\bibfnamefont {E.}~\bibnamefont {Marshman}}\ and\ \bibinfo {author} {\bibfnamefont {C.}~\bibnamefont {Singh}},\ }\bibfield  {title} {\bibinfo {title} {Validation and administration of a conceptual survey on the formalism and postulates of quantum mechanics},\ }\href@noop {} {\bibfield  {journal} {\bibinfo  {journal} {Physical Review Physics Education Research}\ }\textbf {\bibinfo {volume} {15}},\ \bibinfo {pages} {020128} (\bibinfo {year} {2019})}\BibitemShut {NoStop}%
\bibitem [{\citenamefont {Singh}\ and\ \citenamefont {Zhu}(2010)}]{singh2010surveying}%
  \BibitemOpen
  \bibfield  {author} {\bibinfo {author} {\bibfnamefont {C.}~\bibnamefont {Singh}}\ and\ \bibinfo {author} {\bibfnamefont {G.}~\bibnamefont {Zhu}},\ }\bibfield  {title} {\bibinfo {title} {Surveying students’ understanding of quantum mechanics},\ }in\ \href@noop {} {\emph {\bibinfo {booktitle} {AIP Conference Proceedings}}},\ Vol.\ \bibinfo {volume} {1289}\ (\bibinfo {organization} {American Institute of Physics},\ \bibinfo {year} {2010})\ pp.\ \bibinfo {pages} {301--304}\BibitemShut {NoStop}%
\bibitem [{\citenamefont {Goldhaber}\ \emph {et~al.}(2009)\citenamefont {Goldhaber}, \citenamefont {Pollock}, \citenamefont {Dubson}, \citenamefont {Beale},\ and\ \citenamefont {Perkins}}]{goldhaber2009transforming}%
  \BibitemOpen
  \bibfield  {author} {\bibinfo {author} {\bibfnamefont {S.}~\bibnamefont {Goldhaber}}, \bibinfo {author} {\bibfnamefont {S.}~\bibnamefont {Pollock}}, \bibinfo {author} {\bibfnamefont {M.}~\bibnamefont {Dubson}}, \bibinfo {author} {\bibfnamefont {P.}~\bibnamefont {Beale}},\ and\ \bibinfo {author} {\bibfnamefont {K.}~\bibnamefont {Perkins}},\ }\bibfield  {title} {\bibinfo {title} {Transforming upper-division quantum mechanics: Learning goals and assessment},\ }in\ \href@noop {} {\emph {\bibinfo {booktitle} {AIP Conference Proceedings}}},\ Vol.\ \bibinfo {volume} {1179}\ (\bibinfo {organization} {American Institute of Physics},\ \bibinfo {year} {2009})\ pp.\ \bibinfo {pages} {145--148}\BibitemShut {NoStop}%
\bibitem [{\citenamefont {Falk}(2004)}]{falk2004developing}%
  \BibitemOpen
  \bibfield  {author} {\bibinfo {author} {\bibfnamefont {J.}~\bibnamefont {Falk}},\ }\bibfield  {title} {\bibinfo {title} {Developing a quantum mechanics concept inventory},\ }\href@noop {} {\bibfield  {journal} {\bibinfo  {journal} {Unpublished master’s thesis. Uppsala University, Uppsala, Sweden}\ } (\bibinfo {year} {2004})}\BibitemShut {NoStop}%
\bibitem [{\citenamefont {Chasteen}\ \emph {et~al.}(2015)\citenamefont {Chasteen}, \citenamefont {Wilcox}, \citenamefont {Caballero}, \citenamefont {Perkins}, \citenamefont {Pollock},\ and\ \citenamefont {Wieman}}]{chasteen2015educational}%
  \BibitemOpen
  \bibfield  {author} {\bibinfo {author} {\bibfnamefont {S.~V.}\ \bibnamefont {Chasteen}}, \bibinfo {author} {\bibfnamefont {B.}~\bibnamefont {Wilcox}}, \bibinfo {author} {\bibfnamefont {M.~D.}\ \bibnamefont {Caballero}}, \bibinfo {author} {\bibfnamefont {K.~K.}\ \bibnamefont {Perkins}}, \bibinfo {author} {\bibfnamefont {S.~J.}\ \bibnamefont {Pollock}},\ and\ \bibinfo {author} {\bibfnamefont {C.~E.}\ \bibnamefont {Wieman}},\ }\bibfield  {title} {\bibinfo {title} {Educational transformation in upper-division physics: The science education initiative model, outcomes, and lessons learned},\ }\href@noop {} {\bibfield  {journal} {\bibinfo  {journal} {Physical Review Special Topics-Physics Education Research}\ }\textbf {\bibinfo {volume} {11}},\ \bibinfo {pages} {020110} (\bibinfo {year} {2015})}\BibitemShut {NoStop}%
\bibitem [{\citenamefont {Rainey}\ \emph {et~al.}(2020)\citenamefont {Rainey}, \citenamefont {Vignal},\ and\ \citenamefont {Wilcox}}]{rainey2020designing}%
  \BibitemOpen
  \bibfield  {author} {\bibinfo {author} {\bibfnamefont {K.~D.}\ \bibnamefont {Rainey}}, \bibinfo {author} {\bibfnamefont {M.}~\bibnamefont {Vignal}},\ and\ \bibinfo {author} {\bibfnamefont {B.~R.}\ \bibnamefont {Wilcox}},\ }\bibfield  {title} {\bibinfo {title} {Designing upper-division thermal physics assessment items informed by faculty perspectives of key content coverage},\ }\href@noop {} {\bibfield  {journal} {\bibinfo  {journal} {Physical Review Physics Education Research}\ }\textbf {\bibinfo {volume} {16}},\ \bibinfo {pages} {020113} (\bibinfo {year} {2020})}\BibitemShut {NoStop}%
\bibitem [{\citenamefont {Meyer}\ \emph {et~al.}(2024)\citenamefont {Meyer}, \citenamefont {Passante}, \citenamefont {Pollock},\ and\ \citenamefont {Wilcox}}]{meyer2024introductory}%
  \BibitemOpen
  \bibfield  {author} {\bibinfo {author} {\bibfnamefont {J.~C.}\ \bibnamefont {Meyer}}, \bibinfo {author} {\bibfnamefont {G.}~\bibnamefont {Passante}}, \bibinfo {author} {\bibfnamefont {S.~J.}\ \bibnamefont {Pollock}},\ and\ \bibinfo {author} {\bibfnamefont {B.~R.}\ \bibnamefont {Wilcox}},\ }\bibfield  {title} {\bibinfo {title} {Introductory quantum information science coursework at us institutions: content coverage},\ }\href@noop {} {\bibfield  {journal} {\bibinfo  {journal} {EPJ Quantum Technology}\ }\textbf {\bibinfo {volume} {11}},\ \bibinfo {pages} {16} (\bibinfo {year} {2024})}\BibitemShut {NoStop}%
\bibitem [{\citenamefont {Feynman}\ \emph {et~al.}(1965)\citenamefont {Feynman}, \citenamefont {Leighton},\ and\ \citenamefont {Sands}}]{feynman1965lectures3}%
  \BibitemOpen
  \bibfield  {author} {\bibinfo {author} {\bibfnamefont {R.~P.}\ \bibnamefont {Feynman}}, \bibinfo {author} {\bibfnamefont {R.~B.}\ \bibnamefont {Leighton}},\ and\ \bibinfo {author} {\bibfnamefont {M.}~\bibnamefont {Sands}},\ }\href@noop {} {\emph {\bibinfo {title} {The Feynman Lectures on Physics, Volume 3: Quantum Mechanics}}}\ (\bibinfo  {publisher} {Addison-Wesley},\ \bibinfo {address} {Reading, Massachusetts},\ \bibinfo {year} {1965})\BibitemShut {NoStop}%
\bibitem [{\citenamefont {Sakurai}\ and\ \citenamefont {Napolitano}(2020)}]{sakurai2020modern}%
  \BibitemOpen
  \bibfield  {author} {\bibinfo {author} {\bibfnamefont {J.~J.}\ \bibnamefont {Sakurai}}\ and\ \bibinfo {author} {\bibfnamefont {J.}~\bibnamefont {Napolitano}},\ }\href@noop {} {\emph {\bibinfo {title} {Modern quantum mechanics}}}\ (\bibinfo  {publisher} {Cambridge University Press},\ \bibinfo {year} {2020})\BibitemShut {NoStop}%
\bibitem [{\citenamefont {Sadaghiani}(2016)}]{sadaghiani2016spin}%
  \BibitemOpen
  \bibfield  {author} {\bibinfo {author} {\bibfnamefont {H.~R.}\ \bibnamefont {Sadaghiani}},\ }\bibfield  {title} {\bibinfo {title} {Spin first vs. position first instructional approaches to teaching introductory quantum mechanics},\ }in\ \href@noop {} {\emph {\bibinfo {booktitle} {Proc. Phys. Educ. Res. Conf}}}\ (\bibinfo {year} {2016})\ pp.\ \bibinfo {pages} {292--295}\BibitemShut {NoStop}%
\bibitem [{\citenamefont {Riihiluoma}\ \emph {et~al.}(2025)\citenamefont {Riihiluoma}, \citenamefont {Topdemir},\ and\ \citenamefont {Thompson}}]{riihiluoma2025sf}%
  \BibitemOpen
  \bibfield  {author} {\bibinfo {author} {\bibfnamefont {W.~D.}\ \bibnamefont {Riihiluoma}}, \bibinfo {author} {\bibfnamefont {Z.}~\bibnamefont {Topdemir}},\ and\ \bibinfo {author} {\bibfnamefont {J.~R.}\ \bibnamefont {Thompson}},\ }\bibfield  {title} {\bibinfo {title} {Comparative analysis of spins-first and wave functions-first students' understanding of expressions in quantum mechanics},\ }\href {https://doi.org/10.1103/PhysRevPhysEducRes.21.010113} {\bibfield  {journal} {\bibinfo  {journal} {Phys. Rev. Phys. Educ. Res.}\ }\textbf {\bibinfo {volume} {21}},\ \bibinfo {pages} {010113} (\bibinfo {year} {2025})}\BibitemShut {NoStop}%
\bibitem [{\citenamefont {Buzzell}\ \emph {et~al.}(2025)\citenamefont {Buzzell}, \citenamefont {Atherton},\ and\ \citenamefont {Barthelemy}}]{buzzell2025quantum}%
  \BibitemOpen
  \bibfield  {author} {\bibinfo {author} {\bibfnamefont {A.}~\bibnamefont {Buzzell}}, \bibinfo {author} {\bibfnamefont {T.~J.}\ \bibnamefont {Atherton}},\ and\ \bibinfo {author} {\bibfnamefont {R.}~\bibnamefont {Barthelemy}},\ }\bibfield  {title} {\bibinfo {title} {Quantum mechanics curriculum in the us: Quantifying the instructional time, content taught, and paradigms used},\ }\href@noop {} {\bibfield  {journal} {\bibinfo  {journal} {Physical Review Physics Education Research}\ }\textbf {\bibinfo {volume} {21}},\ \bibinfo {pages} {010102} (\bibinfo {year} {2025})}\BibitemShut {NoStop}%
\bibitem [{\citenamefont {{U.S. News \& World Report}}(2023)}]{usnews_physics_rankings_2023}%
  \BibitemOpen
  \bibfield  {author} {\bibinfo {author} {\bibnamefont {{U.S. News \& World Report}}},\ }\href@noop {} {\bibinfo {title} {Best physics schools}} (\bibinfo {year} {2023})\BibitemShut {NoStop}%
\bibitem [{\citenamefont {Singh}\ and\ \citenamefont {Marshman}(2015)}]{singh2015review}%
  \BibitemOpen
  \bibfield  {author} {\bibinfo {author} {\bibfnamefont {C.}~\bibnamefont {Singh}}\ and\ \bibinfo {author} {\bibfnamefont {E.}~\bibnamefont {Marshman}},\ }\bibfield  {title} {\bibinfo {title} {Review of student difficulties in upper-level quantum mechanics},\ }\href@noop {} {\bibfield  {journal} {\bibinfo  {journal} {Physical Review Special Topics—Physics Education Research}\ }\textbf {\bibinfo {volume} {11}},\ \bibinfo {pages} {020117} (\bibinfo {year} {2015})}\BibitemShut {NoStop}%
\bibitem [{\citenamefont {Wittmann}\ \emph {et~al.}(2005)\citenamefont {Wittmann}, \citenamefont {Morgan},\ and\ \citenamefont {Bao}}]{wittmann2005addressing}%
  \BibitemOpen
  \bibfield  {author} {\bibinfo {author} {\bibfnamefont {M.~C.}\ \bibnamefont {Wittmann}}, \bibinfo {author} {\bibfnamefont {J.~T.}\ \bibnamefont {Morgan}},\ and\ \bibinfo {author} {\bibfnamefont {L.}~\bibnamefont {Bao}},\ }\bibfield  {title} {\bibinfo {title} {Addressing student models of energy loss in quantum tunnelling},\ }\href@noop {} {\bibfield  {journal} {\bibinfo  {journal} {European Journal of Physics}\ }\textbf {\bibinfo {volume} {26}},\ \bibinfo {pages} {939} (\bibinfo {year} {2005})}\BibitemShut {NoStop}%
\bibitem [{\citenamefont {Ambrose}(1999)}]{ambrose1999investigation}%
  \BibitemOpen
  \bibfield  {author} {\bibinfo {author} {\bibfnamefont {B.~S.}\ \bibnamefont {Ambrose}},\ }\emph {\bibinfo {title} {Investigation of student understanding of the wave-like properties of light and matter}},\ \href@noop {} {Ph.D. thesis} (\bibinfo {year} {1999})\BibitemShut {NoStop}%
\bibitem [{\citenamefont {Singh}(2008{\natexlab{a}})}]{singh2008student}%
  \BibitemOpen
  \bibfield  {author} {\bibinfo {author} {\bibfnamefont {C.}~\bibnamefont {Singh}},\ }\bibfield  {title} {\bibinfo {title} {Student understanding of quantum mechanics at the beginning of graduate instruction},\ }\href@noop {} {\bibfield  {journal} {\bibinfo  {journal} {American Journal of Physics}\ }\textbf {\bibinfo {volume} {76}},\ \bibinfo {pages} {277} (\bibinfo {year} {2008}{\natexlab{a}})}\BibitemShut {NoStop}%
\bibitem [{\citenamefont {Singh}(2001)}]{singh2001student}%
  \BibitemOpen
  \bibfield  {author} {\bibinfo {author} {\bibfnamefont {C.}~\bibnamefont {Singh}},\ }\bibfield  {title} {\bibinfo {title} {Student understanding of quantum mechanics},\ }\href@noop {} {\bibfield  {journal} {\bibinfo  {journal} {American Journal of Physics}\ }\textbf {\bibinfo {volume} {69}},\ \bibinfo {pages} {885} (\bibinfo {year} {2001})}\BibitemShut {NoStop}%
\bibitem [{\citenamefont {Singh}(2008{\natexlab{b}})}]{singh2008interactive}%
  \BibitemOpen
  \bibfield  {author} {\bibinfo {author} {\bibfnamefont {C.}~\bibnamefont {Singh}},\ }\bibfield  {title} {\bibinfo {title} {Interactive learning tutorials on quantum mechanics},\ }\href@noop {} {\bibfield  {journal} {\bibinfo  {journal} {American Journal of Physics}\ }\textbf {\bibinfo {volume} {76}},\ \bibinfo {pages} {400} (\bibinfo {year} {2008}{\natexlab{b}})}\BibitemShut {NoStop}%
\bibitem [{\citenamefont {Emigh}\ \emph {et~al.}(2020)\citenamefont {Emigh}, \citenamefont {Gire}, \citenamefont {Manogue}, \citenamefont {Passante},\ and\ \citenamefont {Shaffer}}]{emigh2020based}%
  \BibitemOpen
  \bibfield  {author} {\bibinfo {author} {\bibfnamefont {P.~J.}\ \bibnamefont {Emigh}}, \bibinfo {author} {\bibfnamefont {E.}~\bibnamefont {Gire}}, \bibinfo {author} {\bibfnamefont {C.~A.}\ \bibnamefont {Manogue}}, \bibinfo {author} {\bibfnamefont {G.}~\bibnamefont {Passante}},\ and\ \bibinfo {author} {\bibfnamefont {P.~S.}\ \bibnamefont {Shaffer}},\ }\bibfield  {title} {\bibinfo {title} {Research-based quantum instruction: Paradigms and tutorials},\ }\href@noop {} {\bibfield  {journal} {\bibinfo  {journal} {Physical Review Physics Education Research}\ }\textbf {\bibinfo {volume} {16}},\ \bibinfo {pages} {020156} (\bibinfo {year} {2020})}\BibitemShut {NoStop}%
\bibitem [{\citenamefont {Brown}(2015)}]{brown2015developing}%
  \BibitemOpen
  \bibfield  {author} {\bibinfo {author} {\bibfnamefont {B.~R.}\ \bibnamefont {Brown}},\ }\emph {\bibinfo {title} {Developing and assessing research-based tools for teaching quantum mechanics and thermodynamics}},\ \href@noop {} {Ph.D. thesis},\ \bibinfo  {school} {University of Pittsburgh} (\bibinfo {year} {2015})\BibitemShut {NoStop}%
\bibitem [{\citenamefont {Corsiglia}\ \emph {et~al.}(2022)\citenamefont {Corsiglia}, \citenamefont {Pollock},\ and\ \citenamefont {Wilcox}}]{corsiglia_effectiveness_2022}%
  \BibitemOpen
  \bibfield  {author} {\bibinfo {author} {\bibfnamefont {G.}~\bibnamefont {Corsiglia}}, \bibinfo {author} {\bibfnamefont {S.~J.}\ \bibnamefont {Pollock}},\ and\ \bibinfo {author} {\bibfnamefont {B.~R.}\ \bibnamefont {Wilcox}},\ }\bibfield  {title} {\bibinfo {title} {Effectiveness of an online homework tutorial about changing basis in quantum mechanics},\ }in\ \href@noop {} {\emph {\bibinfo {booktitle} {Physics Education Research Conference 2022}}}\ (\bibinfo {year} {2022})\BibitemShut {NoStop}%
\bibitem [{\citenamefont {McKagan}\ \emph {et~al.}(2008)\citenamefont {McKagan}, \citenamefont {Perkins}, \citenamefont {Dubson}, \citenamefont {Malley}, \citenamefont {Reid}, \citenamefont {LeMaster},\ and\ \citenamefont {Wieman}}]{mckagan2008developing}%
  \BibitemOpen
  \bibfield  {author} {\bibinfo {author} {\bibfnamefont {S.}~\bibnamefont {McKagan}}, \bibinfo {author} {\bibfnamefont {K.~K.}\ \bibnamefont {Perkins}}, \bibinfo {author} {\bibfnamefont {M.}~\bibnamefont {Dubson}}, \bibinfo {author} {\bibfnamefont {C.}~\bibnamefont {Malley}}, \bibinfo {author} {\bibfnamefont {S.}~\bibnamefont {Reid}}, \bibinfo {author} {\bibfnamefont {R.}~\bibnamefont {LeMaster}},\ and\ \bibinfo {author} {\bibfnamefont {C.}~\bibnamefont {Wieman}},\ }\bibfield  {title} {\bibinfo {title} {Developing and researching phet simulations for teaching quantum mechanics},\ }\href@noop {} {\bibfield  {journal} {\bibinfo  {journal} {American Journal of Physics}\ }\textbf {\bibinfo {volume} {76}},\ \bibinfo {pages} {406} (\bibinfo {year} {2008})}\BibitemShut {NoStop}%
\bibitem [{\citenamefont {Kohnle}\ and\ \citenamefont {Passante}(2017)}]{kohnle2017sim}%
  \BibitemOpen
  \bibfield  {author} {\bibinfo {author} {\bibfnamefont {A.}~\bibnamefont {Kohnle}}\ and\ \bibinfo {author} {\bibfnamefont {G.}~\bibnamefont {Passante}},\ }\bibfield  {title} {\bibinfo {title} {Characterizing representational learning: A combined simulation and tutorial on perturbation theory},\ }\href {https://doi.org/10.1103/PhysRevPhysEducRes.13.020131} {\bibfield  {journal} {\bibinfo  {journal} {Phys. Rev. Phys. Educ. Res.}\ }\textbf {\bibinfo {volume} {13}},\ \bibinfo {pages} {020131} (\bibinfo {year} {2017})}\BibitemShut {NoStop}%
\bibitem [{\citenamefont {Keebaugh}(2018)}]{keebaugh2018developing}%
  \BibitemOpen
  \bibfield  {author} {\bibinfo {author} {\bibfnamefont {C.~K.}\ \bibnamefont {Keebaugh}},\ }\emph {\bibinfo {title} {Developing and evaluating research-based learning tools for quantum mechanics}},\ \href@noop {} {Ph.D. thesis},\ \bibinfo  {school} {University of Pittsburgh} (\bibinfo {year} {2018})\BibitemShut {NoStop}%
\bibitem [{\citenamefont {{American Council on Education}}(2021)}]{Carnegie:2021}%
  \BibitemOpen
  \bibfield  {author} {\bibinfo {author} {\bibnamefont {{American Council on Education}}},\ }\href {https://carnegieclassifications.acenet.edu/} {\bibinfo {title} {{Carnegie Classification of Institutions of Higher Learning, 2021 ed.}}} (\bibinfo {year} {2021})\BibitemShut {NoStop}%
\bibitem [{\citenamefont {{NASA Minority University Research and Education Project (MUREP)}}(2024)}]{NASA:2024}%
  \BibitemOpen
  \bibfield  {author} {\bibinfo {author} {\bibnamefont {{NASA Minority University Research and Education Project (MUREP)}}},\ }\href {https://msiexchange.nasa.gov/pdf/Final%202024-2025-MSI-List%20(10-23-24).pdf} {\bibinfo {title} {{2024-2025 List of Minority Serving Institutions}}} (\bibinfo {year} {2024})\BibitemShut {NoStop}%
\bibitem [{\citenamefont {Kruse}\ \emph {et~al.}(2024)\citenamefont {Kruse}, \citenamefont {Griston},\ and\ \citenamefont {Wilcox}}]{kruse2024instructors}%
  \BibitemOpen
  \bibfield  {author} {\bibinfo {author} {\bibfnamefont {J.}~\bibnamefont {Kruse}}, \bibinfo {author} {\bibfnamefont {M.}~\bibnamefont {Griston}},\ and\ \bibinfo {author} {\bibfnamefont {B.}~\bibnamefont {Wilcox}},\ }\bibfield  {title} {\bibinfo {title} {Instructors' views on a flexible assessment design},\ }in\ \href {https://doi.org/10.1119/perc.2024.pr.Kruse} {\emph {\bibinfo {booktitle} {Physics Education Research Conference 2024}}}\ (\bibinfo {year} {2024})\ pp.\ \bibinfo {pages} {237--242}\BibitemShut {NoStop}%
\bibitem [{\citenamefont {Henderson}\ and\ \citenamefont {Dancy}(2007)}]{henderson2007barriers}%
  \BibitemOpen
  \bibfield  {author} {\bibinfo {author} {\bibfnamefont {C.}~\bibnamefont {Henderson}}\ and\ \bibinfo {author} {\bibfnamefont {M.~H.}\ \bibnamefont {Dancy}},\ }\bibfield  {title} {\bibinfo {title} {Barriers to the use of research-based instructional strategies: The influence of both individual and situational characteristics},\ }\href@noop {} {\bibfield  {journal} {\bibinfo  {journal} {Physical Review Special Topics—Physics Education Research}\ }\textbf {\bibinfo {volume} {3}},\ \bibinfo {pages} {020102} (\bibinfo {year} {2007})}\BibitemShut {NoStop}%
\bibitem [{\citenamefont {Henderson}\ \emph {et~al.}(2011)\citenamefont {Henderson}, \citenamefont {Beach},\ and\ \citenamefont {Finkelstein}}]{henderson2011facilitating}%
  \BibitemOpen
  \bibfield  {author} {\bibinfo {author} {\bibfnamefont {C.}~\bibnamefont {Henderson}}, \bibinfo {author} {\bibfnamefont {A.}~\bibnamefont {Beach}},\ and\ \bibinfo {author} {\bibfnamefont {N.}~\bibnamefont {Finkelstein}},\ }\bibfield  {title} {\bibinfo {title} {Facilitating change in undergraduate stem instructional practices: An analytic review of the literature},\ }\href@noop {} {\bibfield  {journal} {\bibinfo  {journal} {Journal of research in science teaching}\ }\textbf {\bibinfo {volume} {48}},\ \bibinfo {pages} {952} (\bibinfo {year} {2011})}\BibitemShut {NoStop}%
\end{thebibliography}%

\end{document}